\documentclass[fleqn,usenatbib]{mnras}

\usepackage{newtxtext,newtxmath}
\usepackage[T1]{fontenc}
\usepackage{graphicx}
\usepackage{amsmath}	
\usepackage{hyperref}
\usepackage{threeparttable}
\usepackage{booktabs}
\usepackage{placeins}
\usepackage{pdflscape}
\usepackage{arydshln}
\DeclareRobustCommand{\VAN}[3]{#2}
\let\VANthebibliography\thebibliography
\def\thebibliography{\DeclareRobustCommand{\VAN}[3]{##3}\VANthebibliography}

\defcitealias{Osborn2026}{O26}
\defcitealias{Johnstone2021}{J21}

\title[Neptunian desert variation with stellar mass]{Carving the Neptunian desert as a function of stellar mass and lifetime XUV irradiation}

\author[G. W. King \& A. Osborn et al.]{
George W. King,$^{1,2,3}$\thanks{E-mail: georgekingastro@gmail.com}\footnote[3]{}
Ares Osborn,$^{1,2}$\thanks{Email: dr.ares.osborn@gmail.com}\thanks{These two authors contributed equally to this work and should be considered joint first authors.}
Kaiming Cui,$^{1,2}$
Tim Hallatt,$^{4}$
and David J. Armstrong$^{1,2}$
\\
$^{1}$Centre for Exoplanets and Habitability, University of Warwick, Coventry, CV4 7AL, UK \\
$^{2}$Department of Physics, University of Warwick, Coventry, CV4 7AL, UK \\
$^{3}$Department of Astronomy, University of Michigan, Ann Arbor, MI 48109, USA\\
$^{4}$MIT Kavli Institute for Astrophysics and Space Research, Massachusetts Institute of Technology, Cambridge, MA 02139, USA
}

\date{Accepted XXX. Received YYY; in original form ZZZ}

\pubyear{\the\year{}}

\begin{document}
\label{firstpage}
\pagerange{\pageref{firstpage}--\pageref{lastpage}}
\maketitle

\begin{abstract}
The Neptunian desert is a prominent feature in the exoplanet population, thought to arise from both XUV-driven photoevaporation and tidal circularisation following high-eccentricity migration. However, the extent to which each mechanism contributes, and where, remains uncertain. We examine the desert as a function of stellar type as a test of these mechanisms. 
We determine the desert boundaries for subsamples of F, G, K, and M-hosted planets for radius against period, semi-major axis, bolometric instellation, and lifetime XUV irradiation. Going from F to M, we observe shifts in the boundary between 2.2--7.5\,R$_\oplus$ towards shorter orbital periods, smaller orbital distances, and much smaller instellations. These shifts match changes in lifetime XUV irradiation as a function of stellar mass at fixed period, semi-major axis, and bolometric instellation, evidence towards a photoevaporative origin for the desert at these radii. This is especially the case for orbital period, where the boundary should be invariant across stellar types under the high-eccentricity mechanism.
In lifetime XUV space, we see significant overlap in the boundary across stellar types up to 7.5\,R$_\oplus$. While this could be further evidence towards the XUV-driven mechanism in this regime, our uncertainties cannot rule out a shift in the boundary of the magnitude expected under the high-eccentricity mechanism. 
Overall, our results suggest a photoevaporative origin of the Neptunian desert up to sub-Saturn radii, but increased sample sizes, especially for M-hosted, Neptune-sized planets, will help distinguish further between the mechanisms. 

\end{abstract}




\section{Introduction}

Among the thousands of exoplanets we have discovered over the past few decades, there is a dearth in the number of intermediate-sized planets ($\sim$2-10\,R$_\oplus$) on the shortest orbital periods ($\lesssim$3\,d). Dubbed the ``Neptunian desert'', the observational evidence for this feature stretches back two decades \citep{LDE2007}, and has only become further pronounced as we have continued to discover more planets \citep[e.g.][]{Szabo2011,Beauge2013,Lundkvist2016,Mazeh2016,C-G2024}. More recently, further features in the intermediate-sized planet population have been proposed. A possible overdensity of planets with periods 3-5\,d and radii 5.5-8.5\,R$_{\oplus}$ has been dubbed the ``ridge'' \citep{C-G2024}. This then appears to be followed by another decrease in the occurrence at longer orbital periods, though less sparse than the desert, commonly called the ``savanna'' \citep{Bourrier2023,C-G2024}. The desert in radius-period (and, similarly, mass-period) space cannot be explained by selection effects: we have observed an abundance of smaller planets on similarly short orbital periods, which based on size/mass alone are more difficult signals to detect than planets in the desert would be.

Instead, our current model for the origin of the desert is a combination of two separate physical mechanisms. The first of these is atmospheric escape, driven by X-ray and extreme ultraviolet (EUV; together XUV) photons. This is thought to carve out a large portion of the desert, including setting its lower boundary \citep{Owen2018}. However, photoevaporation alone would lead to an upper boundary at somewhat smaller radii than observed. The upper boundary is instead thought to be the result of the tidal-disruption barrier for large planets moving inwards via high-eccentricity migration \citep[HEM;][]{Matsakos2016,Owen2018,Vissapragada2022}. In this work, we largely focus on XUV-driven escape (but do compare our results to HEM in \S\ref{ssec:HEMcompare}). 

While stars are several orders of magnitude less luminous in XUV than in the optical, XUV photons are absorbed high in the atmosphere of the planet, efficiently heating this region and setting up a hydrodynamic outflow from many close-in planets \citep[e.g.][]{Lammer2003}. Direct observational evidence of atmospheric escape is now plentiful, through the detection of deep transits at several characteristic wavelengths such as Ly\,$\alpha$ \citep[e.g.][]{V-M2003,Ehrenreich2015}, and the near-infrared helium triplet \citep[e.g.][]{Spake2018,DosSantos2023}. Across astronomical timescales, these mass loss processes are capable of driving away substantial fractions of the primordial envelopes of small and intermediate-sized planets, in some cases entirely so \citep[e.g.][]{Owen2012}. In addition to the Neptunian desert \citep[e.g.][]{Owen2018}, this mechanism has also been used to predict and explain the radius valley observed for smaller planets \citep[e.g.][]{Lopez2013,Owen2013,Owen2017}.

The level of XUV emission responsible for driving atmospheric escape changes as a function of stellar age, and is strongly linked to stellar rotation through the latter's role in generating a magnetic dynamo. As stars age, they spin-down due to magnetic braking \citep{Schatzman1962,Skumanich1972}, weakening the dynamo. However, at the earliest stages of a star's life on the main sequence (and therefore in their epoch of fastest rotation), the quiescent ratio of the X-ray and bolometric luminosities, $L_{\rm X}/L_{\rm bol}$, is observed to remain at a constant level of $\sim10^{-3}$ \citep[e.g.][]{Vilhu1987,Stauffer1994,Guedel1997,Jackson2012}. Stars in this regime are commonly referred to as "saturated", in reference to their magnetic dynamos being unable to produce any higher level of emission, even if the star was spun up further.

Eventually, spin-down does begin to reduce the emission level when the star reaches some critical age, which is dependent on several factors including stellar mass and the initial rotation rate. To first-order, the decline in $L_{\rm X}/L_{\rm bol}$ is observed to be a power law, with an index between -1.1 and -1.4 \citep{Ribas2005,Jackson2012}. This steep decline has led to the conclusion that the saturation era is the most important epoch for planetary evolution arising from atmospheric escape \citep[e.g.][]{Lammer2014,Owen2017}. However, EUV time evolution is much less well understood due to the difficulties associated with observing at those wavelengths directly (arising from both strong interstellar absorption and a lack of modern facilities), and a potentially shallower power law may mean that early-ages do not dominate the total EUV irradiation \citep{King2021}.

For individual systems, higher order effects may be important too. Differing initial rotation rates, together with flares, activity cycles, and other variability drive substantial scatter in the observed $L_{\rm X}/L_{\rm bol}$ relationships with both Rossby number \citep[e.g.][]{Pizzolato2003,Wright2011,Wright2018} and age \citep[e.g.][]{Jackson2012,Engle2024,Pezzotti2026}. The extent to which each mechanism contributes to this scatter individually is an outstanding question which requires further work. For flares specifically, several recent studies have suggested that the enhancement in escape due to XUV flaring has only a minor effect on the evolution of Neptune-sized planets and smaller, especially at the close orbital distances of the desert \citep{doAmaral2025,Caldiroli2025}. We note that the impact on Earth-sized planets in the habitable zone around M dwarfs may be much more significant \citep{France2020,doAmaral2025}, but this is not important for this study.

The desert is also not completely empty, with some nomads discovered well within its boundaries \citep{Armstrong2020,Jenkins2020,Osborn2023,Naponiello2023}. Several theories have been put forward to explain their existence. \citet{FernandezFernandez2024} found the desert-planet-hosting star LTT\,9779 to be substantially fainter in X-rays than expected for its age. If this has been true across the system's lifetime, the dimness is sufficient to explain the planet's position in the desert. Alternatively, escape processes in LTT\,9779b and other desert planets could be quenched by their envelopes being metal-rich, which increases cooling \citep{Vissapragada2024,Vissapragada2025}. There are also several mechanisms not involving photoevaporation that could explain desert dwellers, see e.g. \citet{Osborn2023,Hallatt2026,Hallatt2026b,Zanazzi2026}. This remains an active area of research.

The majority of work to date investigating the desert has focussed on FGK stars. Planetary demographics has been dominated by studies of systems observed by \textit{Kepler} \citep{Borucki2010}, whose field of view concentrated on a relatively dim patch of the sky with few nearby targets \citep{Batalha2010}. M dwarf systems were generally more difficult to confirm and characterise due to the lower signal-to-noise \textit{Kepler} obtained for them. In recent years, the Transiting Exoplanet Survey Satellite (\textit{TESS}) has been performing a near-all-sky survey searching for planets \citep{Ricker2015}, which has led to the detection of many nearby planetary systems around M dwarfs which are more amenable to follow up investigations \citep[e.g.][]{Vanderspek2019,Lam2021,Trifonov2021}. Moreover, several spectrographs well-suited for measuring masses of M dwarf-hosted planets, such as CARMENES \citep{Quirrenbach2014}, MAROON-X \citep{Seifahrt2018}, SPIRou \citep{Donati2020}, and NIRPS \citep{Bouchy2025}, have also come online in recent years, further boosting our ability to characterise this population. Recent demographic analyses of FGK-star planetary systems have also begun to leverage \textit{TESS} data \citep[e.g.][]{cuiDemographicsCloseinTESS2026}, complementing the demographic studies from \textit{Kepler}.

Investigating the desert and its boundaries as a function of stellar type, \citet{McDonald2019} concluded that the desert location shifts far more in bolometric instellation space than it does in integrated X-ray flux space. They infer this to be evidence of high-energy emission being the main determinant of the desert location. For the M dwarf desert specifically, assuming mass loss processes dominate, \citet{Hallatt2022} predicted the desert for M dwarfs should be much closer in to the star, with a boundary at a period of ${\sim}$0.7\,d for 0.5\,M$_\odot$ hosts, compared to 3\,d for Solar-like hosts. Recently, \citet{Osborn2026}, hereafter \citetalias{Osborn2026}, compared the desert boundaries for separate M and FGK samples. They found that while the observed M dwarf desert may shift inwards in period space, the change is not as significant as predicted by \citet{Hallatt2022}.

In this study, we extend the work of \citetalias{Osborn2026}. In \S\ref{sec:boundaries}, we update and split up their sample further into separate F, G, K, and M subsamples, and determine observational desert boundaries in orbital period, semi-major axis, and bolometric instellation spaces for each. We then compare the changes in the boundaries across different stellar types to the changes in expected cumulative XUV irradiation as a function of stellar mass in \S\ref{sec:XUVmodel}. In \S\ref{sec:XUVbound}, we combine the XUV models with our observed population to determine the boundaries in lifetime XUV space across stellar types. Finally, we discuss the implications of our findings in \S\ref{sec:discuss}.

\begin{table*}
    \centering
    \caption{Properties of the sample used for the study of the Neptunian desert boundaries, and the subsequent calculated desert opening, as described in \S\ref{sec:boundaries}.}
    \label{tab:sampleprops}
    \begin{threeparttable}
    \begin{tabular}{rcccccc}
    \toprule
                         
    \textbf{Stellar type}           & \textbf{FGKM}     & \textbf{FGK}      & \textbf{F}        & \textbf{G}        & \textbf{K}        & \textbf{M}                 \\
    $T_{\rm eff}$ range (K)         & (7310, 2310]      & (7310, 3890]      & (7310, 5960]      & (5960, 5325]      & (5325, 3890]      & (3890, 2310]      \\
    \midrule
    \multicolumn{7}{c}{Full sample restricted to transit S/N > 7.1} \\
    \midrule
    Num. planets                    & 2791              & 2652              & 897               & 944               & 811               & 139               \\
    Num. stars                      & 2105              & 2004              & 695               & 723               & 586               & 101               \\
    Avg. $M_{\star}$ ($M_{\odot}$)  & $0.96 \pm 0.24$   & $0.99 \pm 0.20$   & $1.20 \pm 0.12$   & $0.98 \pm 0.09$   & $0.76 \pm 0.09$   & $0.42 \pm 0.13$   \\
    \midrule
    \multicolumn{7}{c}{Sample restricted to planets with $1 < R_{\rm P} < 25\,{\rm R_{\oplus}}$,  ${\rm P} < 30$\,d, transit S/N > 7.1} \\
    \midrule
    Num. planets                    & 2193                  & 2074              & 715               & 708               & 651               & 119               \\
    Num. stars                      & 1760                  & 1671              & 590               & 583               & 498               & 89                \\
    Avg. $M_{\star}$ ($M_{\odot}$)  & $0.96 \pm 0.24$       & $0.99 \pm 0.21$   & $1.21 \pm 0.12$   & $0.98 \pm 0.09$   & $0.76 \pm 0.09$   & $0.43 \pm 0.13$   \\
    \hdashline
    \textbf{Desert opening in:} \\
    ... period (d)                  & --                    & $3.2 \pm 0.4$     & $3.8 \pm 0.9$     & $3.5 \pm 0.4$     & $2.9 \pm 0.6$     & $2.2 \pm 0.5$     \\
    ... semi-major axis (au)        & --                    & $0.041 \pm 0.004$ & $0.050 \pm 0.008$ & $0.044 \pm 0.004$ & $0.035 \pm 0.005$ & $0.024 \pm 0.004$ \\
    ... instellation (S$_{\oplus}$) & --                    & $749 \pm 153$     & $969 \pm 323$     & $790 \pm 222$     & $231 \pm 67$      & $48 \pm 17$       \\
    \bottomrule
    \end{tabular}
    \end{threeparttable}
\end{table*}

\section{Building on recent results from Osborn et al. (2026)}
\label{sec:boundaries}

\citetalias{Osborn2026} specifically investigated whether the boundaries of the Neptunian desert differed when considering only planets hosted by M dwarfs, versus those hosted by FGK dwarfs. We summarise the steps they took to do so here, and refer the reader to the paper for the full details. 

To create the sample over which the Neptunian desert boundaries would be explored, \citetalias{Osborn2026} utilised the catalogue of homogeneous, precise stellar parameters for systems hosting planets from \citet{Berger2023}. They restricted it to those with ``confirmed'' and ``known'' planets from Kepler and TESS, and excluded post-main-sequence stellar hosts and those earlier than F0V and later than M9.5V (the latter with a stellar effective temperature cut of $2300 < T_{\rm eff}\,{\rm (K)} < 7300$). \citetalias{Osborn2026} updated the planet radii for the TESS planets using published star-to-planet radius ratios together with the stellar radii from \citet{Berger2023}. They then computed bolometric instellations, $S$, for all planets, to enable a comparison in instellation-radius space in addition to period-radius space. 

Biases in the sample were accounted for following the method of \citet{C-G2024}; in short, this is an inverse detection efficiency method where planets are assigned a ``weight'' based on their detection probability (a combination of geometric transit probability and signal recoverability), with a higher weight given to those with lower detection probability. Planets with a transit S/N < 7.1 were excluded at this stage. M dwarf hosted planets and FGK hosted planets were then split on a $T_{\rm eff}$ of 3900\,K. 

\citetalias{Osborn2026} used a novel method to define the desert boundaries. They defined eight equal size bins in ${\log R_{\rm P}, \log P}$ (or ${\log R_{\rm P}, \log S_{\rm P}}$) space, restricted to $1 < R_{\rm P} < 25\,{\rm R_{\oplus}}$ and ${\rm P} < 30$\,days to avoid uncorrectable incompleteness (this restriction again follows \citealt{C-G2024}). To account for uncertainties in the radii and periods (or instellations) of the planets, they sampled $10^5$ realisations from normal distributions of these based on their reported $1\sigma$ errors. For each of these realisations, the 2.3, 16, 84, and 97.7 percentiles of the orbital periods (or instellations) in each bin were calculated, where each planet was assigned a weight, as explained above, in this calculation. The mean and standard deviation of these percentiles over all realisations were adopted as the final period (or instellation) distributions, and the 2.3 percentile values were taken as tracing the boundary of the desert.
 
For this study, we follow this method, but make a few updates. First, we update again the TESS planet radii with a more recent cross-match to the NASA Exoplanet Archive\footnote{The NASA Exoplanet Archive can be accessed at \url{https://exoplanetarchive.ipac.caltech.edu/}; accessed 12 May 2026.} (using the Planetary Systems Composite Parameters Planet Data table). In addition to radius-period and radius-instellation space, we also look at radius-semi-major axis space (as well as radius-lifetime XUV space in \S\ref{sec:XUVbound}). Following \citet{Pecaut2013}, we define new $T_{\rm eff}$ boundaries for each stellar type, splitting the sample further into F, G, K, and M, as well as the original FGK versus M. The new full sample (prior to the cuts made on transit S/N and in $R_{\rm P}$ and $P$ space) is provided in Table~\ref{tab:sample}, and shown in Fig.~\ref{fig:sample} (here, with the cut to transit S/N applied), and the $T_{\rm eff}$ boundaries and a summary of the properties of the new sample can be found in Table~\ref{tab:sampleprops}. Finally, we reduce the number of realisations sampled from $10^5$ to $10^4$, to decrease computational time (this does not compromise the results). 

\begin{figure*}
    \centering
    \includegraphics[width=\linewidth]{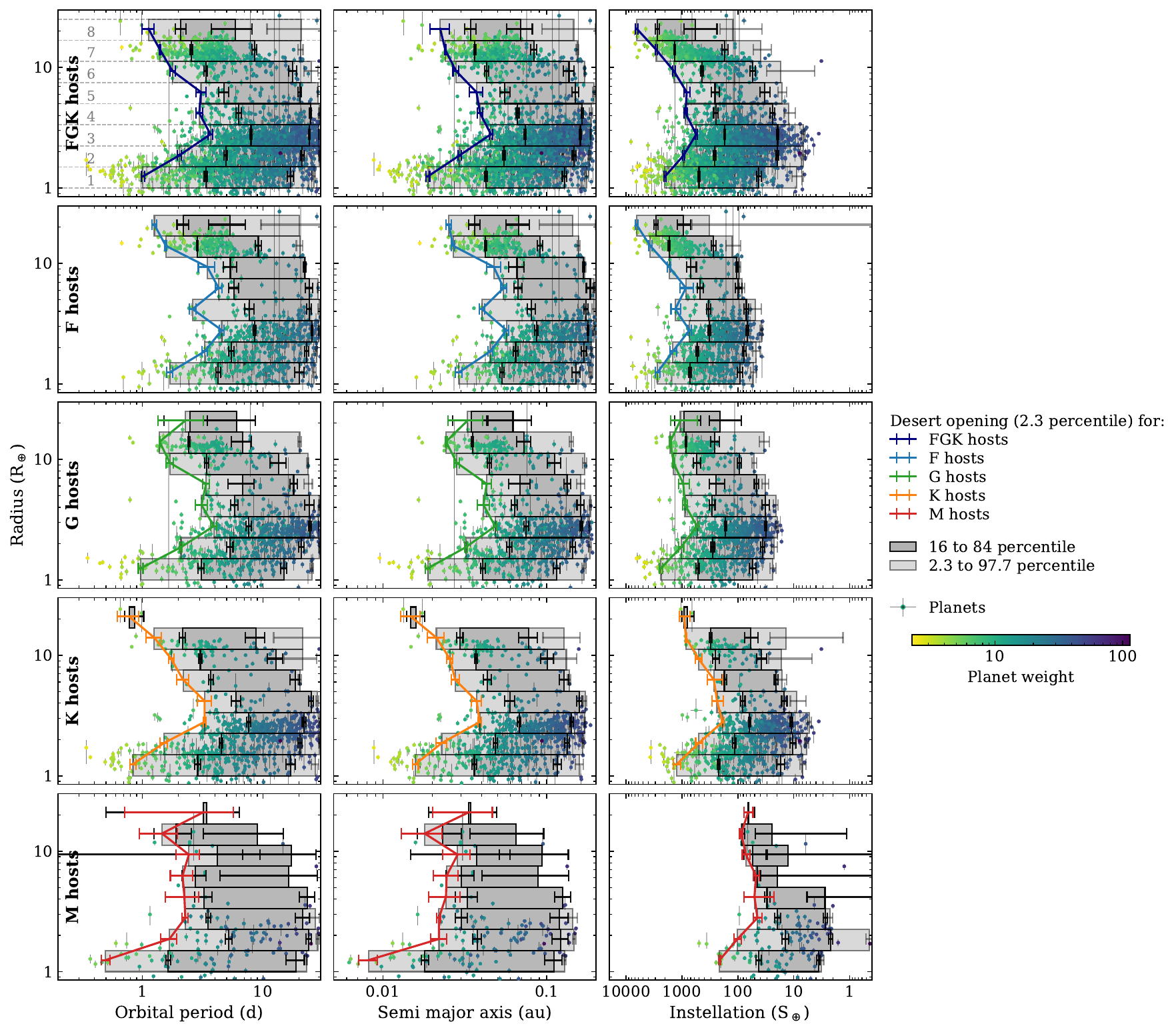}
    \caption{The Neptunian desert boundaries, as described in \S\ref{sec:boundaries}, for our FGK (first row), F (second row), G (third row), K (fourth row), and M dwarf planet hosts (fifth row). These are each shown in radius-period space (left column), radius-semi-major axis space (middle column), and radius-instellation space (right column). Our planet sample (summarised in Table~\ref{tab:sampleprops}, and given in full in Table~\ref{tab:sample}, Fig.~\ref{fig:sample}) is shown as coloured dots, according to the weighting of each planet. The boundaries of the bins in radius space are shown in the top left panel, delineated with dashed lines and numbered from 1-8. For each of our period, semi-major axis, and instellation spaces, the percentile values calculated for each radius bin and their errors are shown, where the 2.3-97.7 percentile range are shaded in light grey, and the 16-84 percentile range is shaded in dark grey. The 2.3 percentile, which we take as the boundary of the Neptunian desert, is traced in colour in each panel.}
    \label{fig:sample-percentiles}
\end{figure*}

\begin{figure*}
    \centering
    \includegraphics[width=\linewidth]{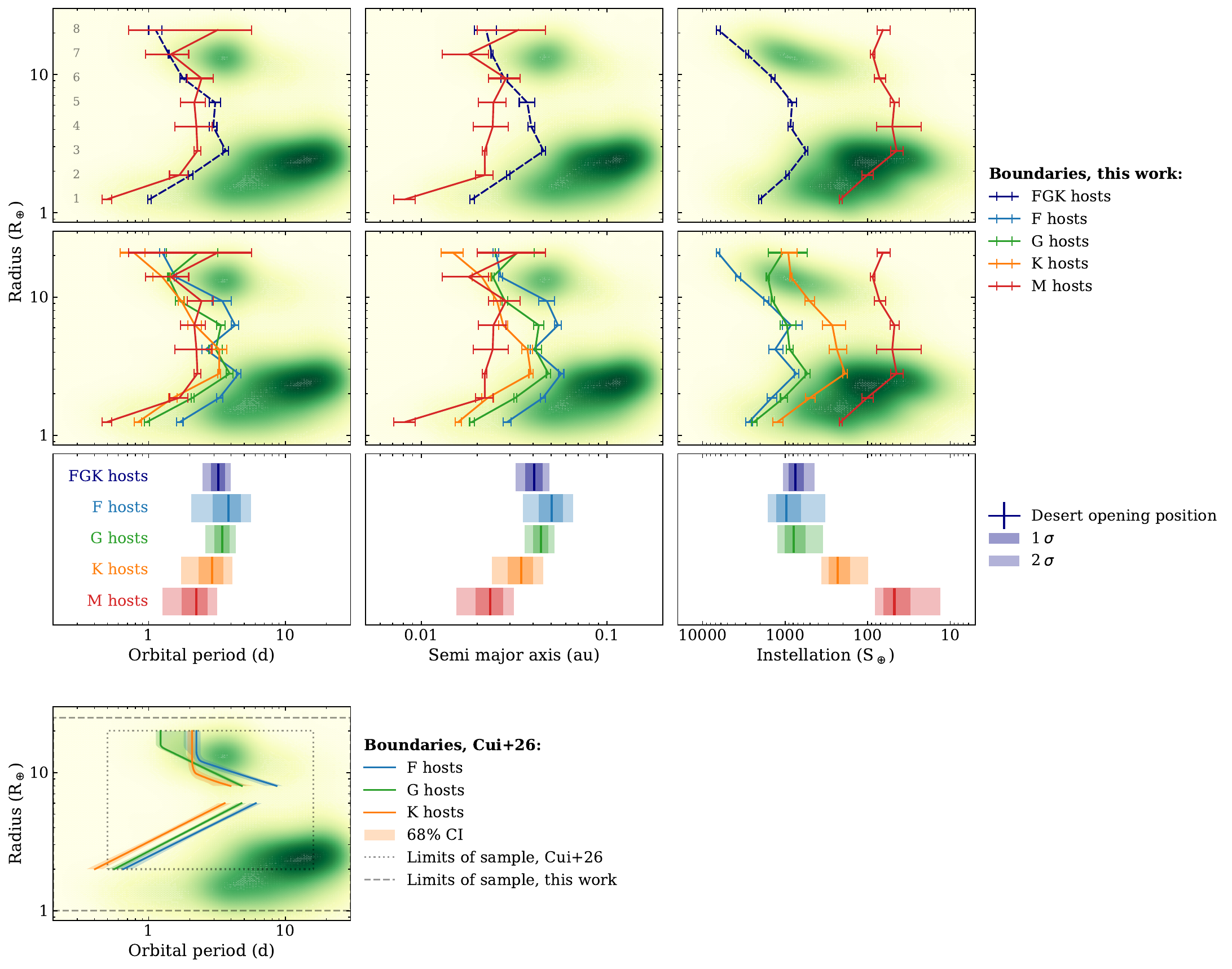}
    \caption{The 2.3 percentile Neptunian desert boundaries are compared between host star spectral types in radius-period space (first column), radius-semi-major axis space (second column), and radius-instellation space (third column). The bins are labelled in the top left panel for reference. The first row simply compares FGK hosts (dark blue) to M dwarf hosts (red) as in \citetalias{Osborn2026}, whereas the second row compares between F (blue), G (green), K (yellow), and M (red) hosts. The contours show the density of our full planet sample (given in Table~\ref{tab:sample}). The third row displays how the desert opening (the average of the 2.3 percentile across bins 3-5) changes between host type, with 2 standard deviations of the values shaded. The final row shows the Neptunian desert boundaries determined by \citet{Cui2026spectype} in period-radius space for F, G, and K hosts, with the 68 per cent confidence intervals shaded. The extent of the parameter space utilised for their boundary determination is shown as the dotted box, contrasted to that used in this work shown as the dashed box.}
    \label{fig:sample-lines}
\end{figure*}

We show the 2.3, 16, 84, and 97.7 percentiles of radius-period, -semi-major axis, and -instellation bins for the FGK, F, G, K, and M samples in Fig.~\ref{fig:sample-percentiles}, and overplot the 2.3 percentile desert boundary. We take these boundaries and compare between stellar types more directly for each parameter space in Fig.~\ref{fig:sample-lines}. Following \citetalias{Osborn2026}, we define the ``opening'' of the desert to cover bins 3-5 i.e., from 2.2 to 7.5\,R$_{\oplus}$ (the bins are labelled in Fig.~\ref{fig:sample-percentiles}, going from bin 1 at the lowest radii to bin 8 at the highest radii). This is a radius range where the upper and lower boundaries intersect and the relative contributions of photoevaporation and HEM remain most uncertain. The period (or instellation) value of the opening was calculated as a weighted mean of the 2.3 percentile values for bins 3-5 in \citetalias{Osborn2026}, but we discovered a mistake in the calculation of the error on the weighted mean. Here, we instead do the calculation by taking the $10^4$ values of the 2.3 percentile from each of bins 3-5, and then taking the mean and standard deviation of all of these. The desert opening values in period, semi-major axis, and instellation, across each of our stellar type samples, are presented in Table~\ref{tab:sampleprops} and are displayed in the third row of Fig.~\ref{fig:sample-lines}. 

The desert opening values in period from \citetalias{Osborn2026} for the FGK versus M samples were $3.3 \pm 1.4$\,d and $2.2 \pm 1.0$\,d, respectively, which match the values here but our new errors are a factor of 2--3 smaller (due to the previously incorrect calculation). So, the desert opening shifts about $1.6\sigma$ in period (compared to $0.6\sigma$ presented previously). The conclusion remains the same: the opening of the desert for M dwarf hosted planets still shifts slightly inwards in period in comparison to FGK hosted planets, though not as significantly as predicted by \citet{Hallatt2022}. We can also now see this inward shift happen gradually as we move from F hosts to M hosts. Additionally, the shift is evident for the lower boundary at lower radii, though becomes muddled at the upper boundary at higher radii -- for the M host sample in particular, the very small sample size of giant planets is a limiting factor here. In instellation, the values from \citetalias{Osborn2026} for the FGK versus M samples were $830 \pm 0.1\,S_{\oplus}$ and $57.8 \pm 0.2\,S_{\oplus}$. Again, these values are similar to our new values presented here, but the errors here are largely increased (again due to the previously incorrect calculation). Here, we see a significant shift of $4.6\sigma$. Our conclusion still remains the same: there is a significant shift in the desert opening to lower instellation when going from FGK to M hosts. This can also be seen to happen gradually as we move from F hosts to M hosts, and for the lower boundary too, but not so much the upper. Finally, in the new parameter space of semi-major axis (not previously explored in \citetalias{Osborn2026}), the boundary shifts to lower semi-major axes for M dwarfs compared to FGKs. This shift, a $3\sigma$ difference, is a bit more significant than in period, but not as much so as instellation. Once again, this shift happens gradually as we move from F hosts to M hosts. Here it is also evident for the upper boundary as well as the lower boundary.

Recently, \citet{Cui2026spectype} also investigated the Neptunian desert boundaries with different stellar samples in period-radius space. The boundaries were fit from an occurrence rate distribution based on TESS discoveries only, and it was split into F, G, and K type hosts, not extending down to M dwarfs. Despite these differences, we still see the same behaviour happening with the boundaries, shown in Fig.~\ref{fig:sample-lines}: the lower boundary and the opening of the desert in particular shift to lower orbital periods as we go from F to K hosts.

\section{Lifetime irradiation modelling}
\label{sec:XUVmodel}

We next aim to explore whether the observed desert boundary changes between stellar types can be explained by differences in the XUV irradiation budget as a function of stellar mass.

\subsection{Modelling setup}

For our lifetime XUV model, we adopt the \textsc{Mors} code, detailed in \citet{Johnstone2021}, hereafter \citetalias{Johnstone2021}. \textsc{Mors} is a suite of semi-empirical evolutionary tracks that map out the X-ray, EUV, and Ly\,$\alpha$ luminosities, as well as the rotation of stars, all as a function of both age and mass. The tracks also model the effects of different initial rotation rates, which can have a substantial impact on the high-energy emission of a star of given mass between about 10\,Myr and a few Gyr, with the exact range of divergence depending on stellar mass \citep[e.g.][\citetalias{Johnstone2021}]{Tu2015}. In its use of observational constraints on XUV emission, \citetalias{Johnstone2021} did not distinguish flaring epochs from quiescence, meaning stellar variability due to this, as well as other effects such as activity cycles, are implicitly accounted for in the \textsc{Mors} tracks too. 

For their extrapolation from X-ray to the EUV, \citetalias{Johnstone2021} derives a scaling of the surface fluxes in the two bands. They motivate this choice by noting that surface flux shows a mass-independent correlation with the mean plasma temperature (weighted by the respective emission measures), while neither $L_{\rm X}$ or $L_{\rm X}/L_{\rm bol}$ do. They base their relation on a combination of \textit{Extreme Ultraviolet Explorer} (\textit{EUVE}) observations and the Sun, describing their methods in detail in their section 4.1. When converted to a temporal evolution, their extrapolation leads to a slightly shallower drop off in EUV emission than \citet{Sanz-Forcada2011}, but slightly steeper than \citet{King2021}.

We used \textsc{Mors} to generate evolutionary tracks up to 5\,Gyr for stellar masses between 0.1 and 1.2\,M$_\odot$, with a step size of 0.01\,M$_\odot$. This upper mass limit set by \textsc{Mors} cuts early-to-mid F stars from consideration. We note that we tested the effect on the fitted boundaries from \S\ref{sec:boundaries} of limiting the F and FGK subsamples to only host stars with masses $\leq1.2$\,M$_\odot$. We found the results of every radius bin in all three parameter spaces agreed with its position as calculated in \S\ref{sec:boundaries} within 1-$\sigma$. Thus, \textsc{Mors} only being able to calculate tracks for stars $\leq1.2$\,M$_\odot$ does not affect the comparisons we make in \S\ref{ssec:compBoundwMors}.

We calculated tracks at each mass step for the median initial rotational rate, as well the 2.3 and 97.7 percentiles, to obtain $\pm2\sigma$ confidence regions. The $L_{\rm bol}$ track generated by \textsc{Mors} is always the same for a given stellar mass. These values are taken from the earlier stellar evolution models of \citet{Spada2013}, and are not modelled with any dependence on initial rotation rate. Thus, our error regions are solely reflective of the high-energy and rotation properties, although these should be expected to dominate the uncertainties anyway given the $\sim$order of magnitude scatter in observed high-energy relationships with rotation and age \citep[e.g.][]{Wright2011,Jackson2012}.

\subsection{Results}
\label{ssec:XUVmodRes}

\begin{figure}
    \centering
    \includegraphics[width=\columnwidth]{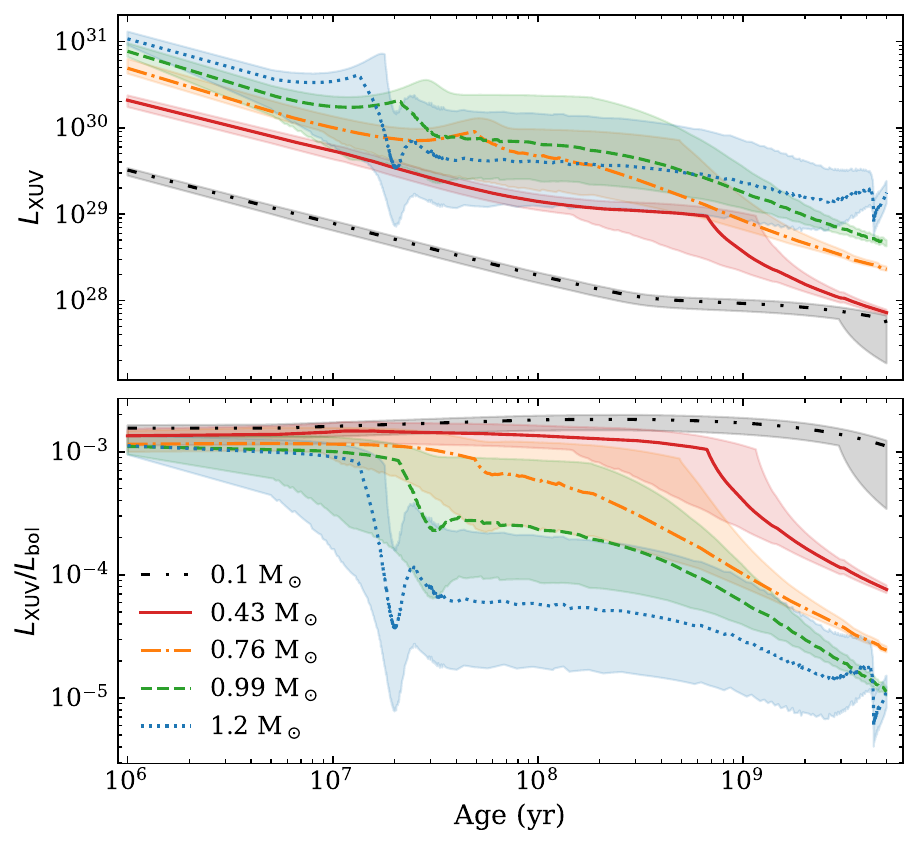}
    \caption{Examples of the evolution of $L_{\rm XUV}$ and $L_{\rm XUV}/L_{\rm bol}$ with age for a few different stellar masses. In each case, we plot the track corresponding to the median initial rotation rate. The shaded region represents the 2-$\sigma$ uncertainty, i.e. the lower and upper limits of the shaded regions are the tracks corresponding to the 2.3 and 97.7 percentiles of the initial rotation rate}.
    \label{fig:Lxuv+ratios}
\end{figure}



\begin{figure}
    \centering
    \includegraphics[width=\columnwidth]{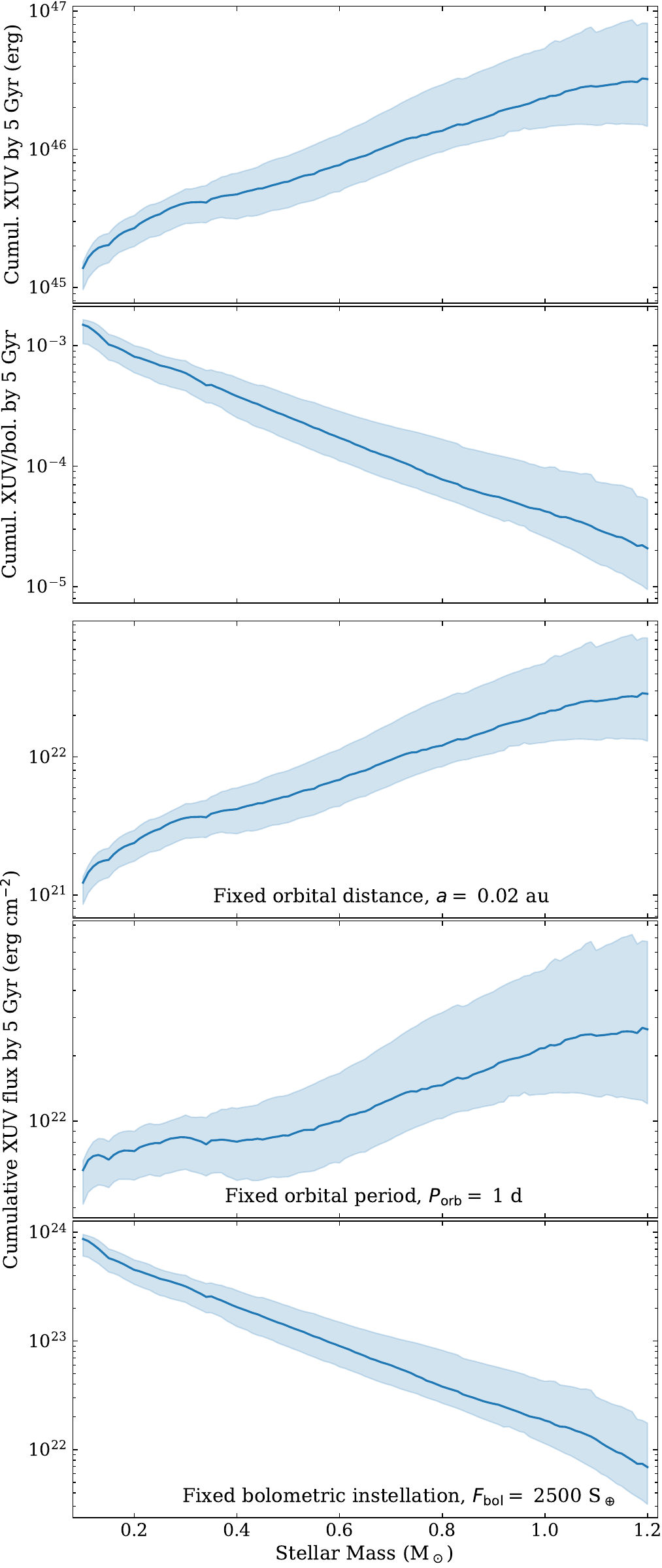}
    \caption{Various plots of cumulative XUV as a function of stellar mass. The top two panels plot cumulative XUV energy emitted up to 5\,Gyr in absolute terms (top panel) and as a fraction of the cumulative bolometric energy emitted (second panel). The bottom three panels plot cumulative XUV irradiation flux up to 5\,Gyr as a function of stellar mass at fixed orbital distance (third panel), orbital period (fourth panel), and bolometric instellation (bottom panel). The shaded region represents the 2-$\sigma$ uncertainty.}
    \label{fig:cumulComb}
\end{figure}

\begin{table*}
\centering
\caption{Comparison of the cumulative luminosities and fluxes up to 5\,Gyr for the average stellar mass in the M dwarf (0.43\,M$_\odot$) and G (0.98\,M$_\odot$) samples. All uncertainties are 2-$\sigma$ and based on the rotation rate distribution in \citetalias{Johnstone2021}.}
\label{tab:M+FGKcomp}
\begin{tabular}{cccc}
\hline
Parameter (Cumulative to 5\,Gyr)   & 0.43\,M$_\odot$ value                                & 0.98\,M$_\odot$ value                              & Unit  
\\ \hline
X-ray Energy                       & $\left( 2.2^{+1.2}_{-0.9} \right) \times 10^{45}$    & $\left( 9.2^{+18.4}_{-4.3} \right) \times 10^{45}$ & erg            \\
EUV Energy                         & $\left( 2.9^{+1.0}_{-0.8} \right) \times 10^{45}$    & $\left( 1.3^{+1.1}_{-0.4} \right) \times 10^{46}$  & erg            \\
XUV Energy                         & $\left( 5.1^{+2.2}_{-1.8} \right) \times 10^{45}$    & $\left( 2.2^{+2.9}_{-0.8} \right) \times 10^{46}$  & erg            \\
Bolometric Energy                        & $1.5 \times 10^{49}$                                 & $5.0 \times 10^{50}$                               & erg            \\
Cumul. XUV/ Cumul. Bolometric                     & $\left( 3.4^{+1.4}_{-1.2} \right) \times 10^{-4}$    & $\left( 4.4^{+5.8}_{-1.6} \right) \times 10^{-5}$  & -              \\
XUV flux for 0.02\,au orbit        & $\left( 4.5^{+1.9}_{-1.6} \right) \times 10^{21}$    & $\left( 2.0^{+2.6}_{-0.7} \right) \times 10^{22}$  & erg\,cm$^{-2}$ \\
XUV flux for 1\,d orbit            & $\left( 8.2^{+3.5}_{-2.9} \right) \times 10^{21}$    & $\left( 2.1^{+2.7}_{-0.8} \right) \times 10^{22}$  & erg\,cm$^{-2}$ \\
XUV flux at 2500\,S$_\oplus$ orbit & $\left( 1.81^{+0.78}_{-0.64} \right) \times 10^{23}$ & $\left( 2.0^{+2.6}_{-0.7} \right) \times 10^{22}$  & erg\,cm$^{-2}$ \\ \hline
\end{tabular}
\end{table*}

In Fig.~\ref{fig:Lxuv+ratios}, we plot examples of the XUV luminosity evolution as a function of age, for both absolute energy units and with respect to $L_{\rm bol}$. Two of the curves plotted are for the average stellar masses of the M and combined FGK samples in \S\ref{sec:boundaries}, while a further two are the lower and upper ends of the mass range considered (0.1 and 1.2\,M$\odot$). The other is for the average mass of the K star sample, since the others plotted roughly correspond to late F, G-type, early-mid M, and late M.

Throughout this section we plot the results of our investigations of lifetime irradiation for the combined XUV energy range. While the values of this and the separate X-ray and EUV components differ in each case, the morphological differences in the curves we plot as a function of age/mass between the energy bands are rather subtle. As a result, the overall conclusions remain the same if just the X-ray or EUV alone are plotted. For comparison with Fig.~\ref{fig:Lxuv+ratios}, we provide separate example X-ray and EUV time evolution plots in Appendix~\ref{app:timeEvo}.

The bottom panel of Fig.~\ref{fig:Lxuv+ratios} demonstrates several key findings of X-ray (and thus XUV) time evolution studies for different stellar types over the past few decades. Firstly, when scaled by $L_{\rm bol}$, all stellar masses initially cluster around $\sim 10^{-3}$, in line with the observed saturation at the youngest ages. Secondly, the length of the saturation period is dependent on stellar mass, with less massive stars remaining saturated for longer. Thirdly, the observed drop off rate after saturation is similar across many stellar types (e.g. compare the 0.43, 0.76, and 0.99\,M$_\odot$ tracks at $>1$\,Gyr).

Consequently, less massive stars generally output a greater fraction of their bolometric emission as XUV after $\sim 10$\,Myr than their higher mass counterparts. However, this is the opposite of what happens when considering absolute energy units, as shown in the top panel. Here, the emitted energy per unit time is generally higher for more massive stars across all ages. As \citetalias{Johnstone2021} also points out in their discussion, this is an extra nuance that is not encapsulated by the simplistic description, common in the exoplanet literature, of M dwarfs being more active than Solar-like stars. While true as a fraction of their bolometric flux, it does not mean M dwarfs are more XUV luminous.

To investigate further, we now consider cumulative XUV emission and fluxes as a function of stellar mass. In the top two panels of Fig.~\ref{fig:cumulComb}, we plot the cumulative XUV energy emitted by stars up to a typical field age of 5\,Gyr, across our stellar mass range. This plot demonstrates even more clearly what we saw in Fig.~\ref{fig:Lxuv+ratios}: more massive stars emit more total XUV energy than less massive stars, but this is a lower fraction of the cumulative bolometric emission. 

For the desert and its boundaries, the combined effects of these two key points depends on the parameter space in which you plot the exoplanet population. In the bottom three panels of Fig.~\ref{fig:cumulComb}, we plot the cumulative XUV flux received by a planet up to 5 Gyr as a function of stellar mass for three cases: 1) at a fixed orbital distance (0.02\,au); 2) at a fixed orbital period (1\,d); 3) at a fixed bolometric instellation (2500\,S$_\oplus$).

Comparing the different plots within Figure \ref{fig:cumulComb}, we see that the shape of the curves in the top and third panels of match one another closely. For a fixed orbital distance, $a$, the cumulative flux is simply the cumulative luminosity divided by $4\pi a^2$. Since $a$ is the same across all stellar masses in this instance, the division is the same, and so we reproduce exactly the same curve from the top panel in the third panel, just scaled by the $4\pi a^2$ factor. 

For the fixed instellation case, the orbital distance we scale to is set by the \textit{instantaneous} $L_{\rm bol}$ at 5\,Gyr. However, while $L_{\rm bol}$ does change slightly over the system lifetime up to this age, changes after reaching the main sequence are small, particularly in comparison to the changes in XUV. As such, if one multiplies the $L_{\rm bol}$ at 5\,Gyr by that system age in seconds (i.e. assuming that instantaneous level has not changed over the system lifetime), this deviates from the cumulative energy of its \textsc{Mors} bolometric track by less than a factor of two for all stellar masses considered. Consequently, the relative scaling as a function of stellar mass in the bottom panel of Fig.~\ref{fig:cumulComb} is very similar to the cumulative bolometric scaling in the second panel, and so the curve shapes are largely indistinguishable.

The fourth panel of Fig.~\ref{fig:cumulComb} for a fixed orbital period shows some similarity to the top and third panels, but the increase towards higher mass stars is more gentle. The orbital distance corresponding to a 1\,d period increases slightly with stellar mass (since for a Keplerian orbit at fixed orbital period $a_{\rm fix P}\propto M_{\rm*}^\frac{1}{3}$), acting to somewhat moderate the increase in cumulative XUV when dividing the curve in the top panel by those distances.

\subsection{Comparison with the observed desert boundaries}
\label{ssec:compBoundwMors}

If XUV-driven escape shapes the desert and sets a substantial portion of its boundaries, we should expect similar lifetime XUV fluxes at the location of the boundary. Ahead of the following comparisons, we remind the reader that cutting the sample to only those hosted by stars $\leq1.2$\,M$_\odot$ to match \textsc{Mors} does not affect the observed boundaries within the uncertainties.

We can see that the shifts in boundary location in Fig.~\ref{fig:sample-lines} match inversely the variations in cumulative XUV flux as a function of stellar mass at fixed values of the same parameters in Fig.~\ref{fig:cumulComb}. In semi-major axis space, the best fit boundary shifts inwards to smaller orbital distances with decreasing stellar mass. To get an equivalent dose of cumulative XUV, a planet must be closer to an M star than a Solar-like star. This is because at a specific orbital distance around the M star, compared to the same distance from a Solar-like star, the integrated irradiation is lower. The smaller inward shifts seen in orbital period space compared to semi-major axis space, both in our study and \citet{Cui2026spectype}, mirrors the more gentle slope observed in the cumulative XUV flux at fixed orbital period. Finally, the steep negative slope in cumulative XUV at fixed $F_{\rm bol}$ also matches the significant shift in the desert boundary to lower instellations for lower stellar masses.

For a further, more quantitative comparison, in Table~\ref{tab:M+FGKcomp} we list various cumulative luminosities and fluxes for two particular stellar masses: one for the average mass in the M dwarf (0.43\,M$_\odot$) sample, and one for the average mass in the G (0.98\,M$_\odot$) sample. The latter is also essentially the average mass of the FGK-only sample too. Notably, we see that up to an age of 5\,Gyr, the typical Solar-like star emits a little over four times more XUV than the M dwarf. Accordingly, the shift of the desert opening in semi-major axis space listed in Table~\ref{tab:sampleprops} is just less than a factor of two, roughly matching the factor change in XUV since flux is inversely proportional to the square of the orbital distance.

\section{Boundaries in lifetime XUV irradiation space}
\label{sec:XUVbound}

We now combine our work from the previous two sections together. We aim to analyse the Neptunian desert boundary in the lifetime XUV parameter space, and directly test the statement we made at the top of \S\ref{ssec:compBoundwMors}: if XUV-driven escape is carving a significant portion of the boundary, it is a more natural space in which to investigate the desert, and the boundaries should not differ across stellar types. Of course, the trade off is that this is a much harder quantity to constrain than any of the other three variables considered in \S\ref{sec:boundaries}. 

To determine the boundaries in this space, we use the same sample from \S\ref{sec:boundaries}. We aim to simulate the lifetime XUV experienced by these planets at a population level, and then use that to assess the desert. We started by generating a new, much larger set of lifetime tracks with \textsc{Mors}. We retained the same mass limits (0.1 to 1.2\,M$_\odot$) and mass step size (0.01\,M$_\odot$) as before, but this time generated tracks at every integer percentile in initial rotation rate from 0 to 100, inclusive. Based on the stellar mass restrictions of \textsc{Mors}, we made an additional cut to the sample from \S\ref{sec:boundaries} and detailed in Table~\ref{tab:sampleprops}: all stars whose best fit mass was in excess of 1.2\,M$_\odot$ had their system excluded from this analysis. The properties of this sample are listed in Table~\ref{tab:xuvsampleprops}, with the biggest change being to the F stars, whose sample is roughly halved. A handful of the G sample are excluded, and unsurprisingly the K and M samples are unaffected. The resulting average masses of the F and combined FGK samples are marginally lower.

We then employed a similar bootstrapping technique to \citetalias{Osborn2026} and \S\ref{sec:boundaries} over the ranges of radii and cumulative XUV each planet could have. Each of the realisations of cumulative XUV for each planet required a multi-step process to generate them, as this is not a directly observable quantity. For each system at each realisation, we drew a value of stellar mass and rounded to two decimal places to match the step size in our generated tracks. We assumed a Gaussian distribution where the measured errorbars were symmetric. Where not, we adopted the same policy as in \citetalias{Osborn2026} (and thus implicitly in \S\ref{sec:boundaries}): we piecemealed together two Gaussians with standard deviations equal to the the upper and lower uncertainties on their respective sides of the best fit value. Note that we also truncated any distribution that went above 1.205\,M$_\odot$, such that no values could be drawn above 1.2\,M$_\odot$ after rounding. We then drew a random integer value for the initial rotation percentile between 0 and 100, inclusive, and found the track corresponding to the drawn rounded mass and percentile.

To find the cumulative XUV energy from the track we needed to draw a value for the stellar age, with the values again taken from \citet{Berger2023}. We applied a similar (piecemeal, where necessary) Gaussian distribution approach using the best fit age and uncertainties as above. In this instance, we truncated the distributions at an age of zero and the maximum age in the \textsc{Mors} track, as necessary. The maximum age \textsc{Mors} calculates a track to is either the point of evolution off the main sequence at the high mass end, or 13\,Gyr at the low mass end - roughly matching the age of the Universe. We then linearly interpolated the Mors track to find the cumulative XUV according at that age.

We note the quoted stellar age for individual systems in the field above $\sim$1\,Gyr often have large (100 per cent or more) uncertainties. However, we are helped by the fact that by a few Gyr in age, the decline in high-energy emission is such that the increases in cumulative XUV are relatively slow. For example, the cumulative XUV energy output by a 1\,M$_\odot$ star with median initial rotation only increases by a factor of two from 1.3 to 10\,Gyr (with a similar 1.6 factor difference across the same age range for a 0.4\,M$_\odot$ star too). As such, comparing with Fig.~\ref{fig:cumulComb}, the cumulative XUV of stars above a few Gyr old is far more sensitive to the stellar mass than to the specific age within that range. We note that future studies of X-ray decline with more precise ages (e.g. from PLATO) could demonstrate significant differences from the rate of decline used by \citetalias{Johnstone2021} in developing Mors, but the deviation would need to be substantial to match that from stellar mass across a population.

Our final draws for each realisation were from the distributions of possible semi-major axis (to scale the cumulative energy to a cumulative irradiation per unit area), and planetary radii (for the y-axis) values. We sample 25,000 realisations, higher than before due to sampling over several parameters to produce a cumulative XUV. We then applied the same process of calculating percentiles within the different radius bins, adopting the mean (along with the standard deviation for the errors) of the upper 2-$\sigma$ percentile draws as the boundary of the desert.

\begin{table*}
    \centering
    \caption{Properties of the sample used for the study of the Neptunian desert boundaries in lifetime XUV space, and the subsequent calculated desert opening.}
    \label{tab:xuvsampleprops}
    \begin{tabular}{rcccccc}
    \toprule
                         
    \textbf{Stellar type}           & \textbf{FGKM}     & \textbf{FGK}      & \textbf{F}        & \textbf{G}        & \textbf{K}        & \textbf{M}        \\
    $T_{\rm eff}$ range (K)         & (7310, 2310]      & (7310, 3890]      & (7310, 5960]      & (5960, 5325]      & (5325, 3890]      & (3890, 2310]      \\
    \midrule
    \multicolumn{7}{c}{Sample restricted to planets with $1 < R_{\rm P} < 25\,{\rm R_{\oplus}}$,  ${\rm P} < 30$\,d, transit S/N > 7.1; and stars with $M_* <= 1.2$\,M$_\odot$} \\
    \midrule
    Num. planets                    & 1894                  & 1775              & 423               & 701               & 651               & 119               \\
    Num. stars                      & 1502                  & 1413              & 339               & 576               & 498               & 89                \\
    Avg. $M_{\star}$ ($M_{\odot}$)  & $0.90 \pm 0.20$       & $0.93 \pm 0.16$   & $1.12 \pm 0.06$   & $0.98 \pm 0.08$   & $0.76 \pm 0.09$   & $0.43 \pm 0.13$   \\
    \hdashline
    \textbf{Desert opening in:} \\
    ... lifetime XUV ($10^{21}$\,erg\,s$^{-2}$)    & --                    & $5.3 \pm 1.3$     & $4.8 \pm 2.3$     & $6.1 \pm 2.1$     & $4.7 \pm 1.6$     & $3.7 \pm 1.6$     \\
    \bottomrule
    \end{tabular}
\end{table*}

\subsection{Results}
\label{ssec:xuvResults}

\begin{figure}
    \centering
    \includegraphics[width=\columnwidth]{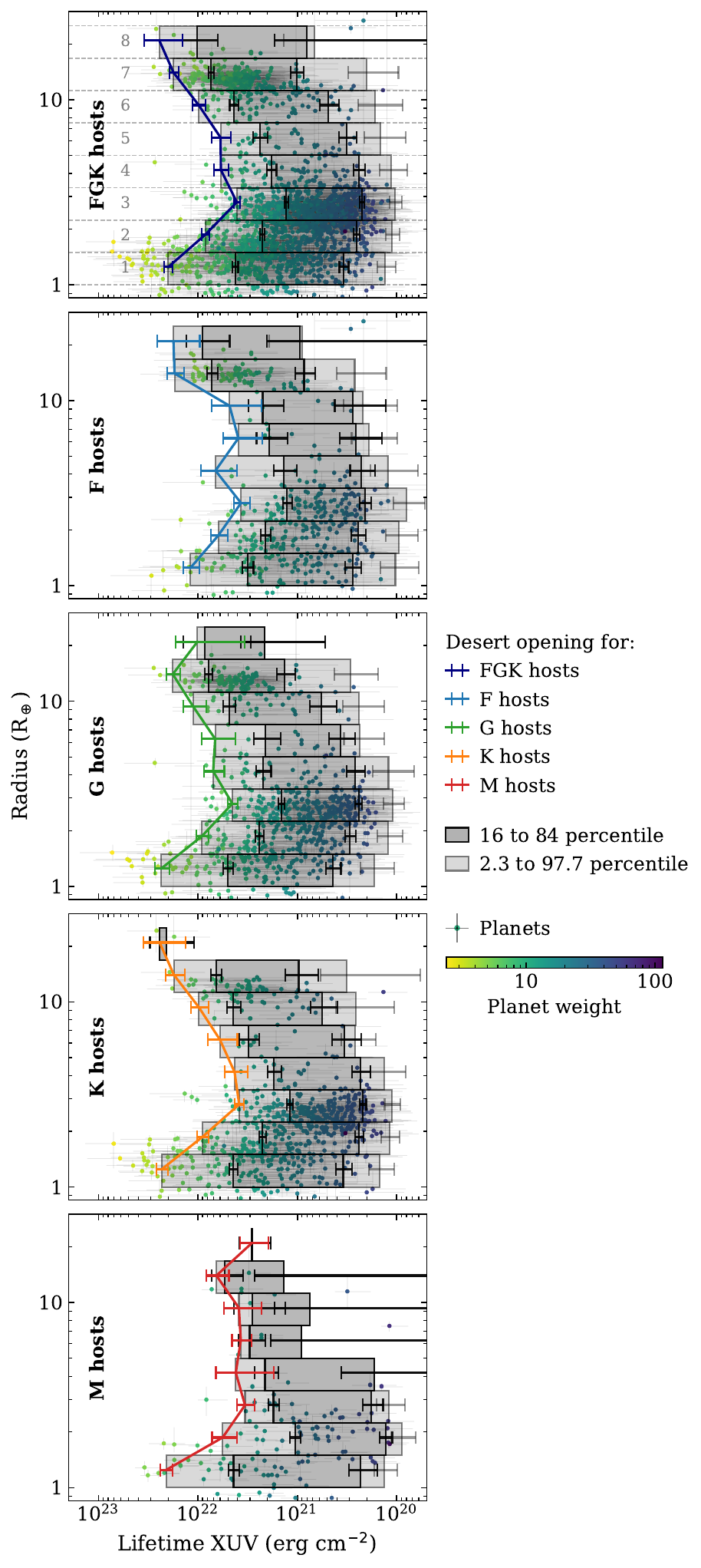}
    \caption{Similar to Fig.~\ref{fig:sample-percentiles}, but in lifetime XUV space.}
    \label{fig:XUVboundFit}
\end{figure}

\begin{figure}
    \centering
    \includegraphics[width=\columnwidth]{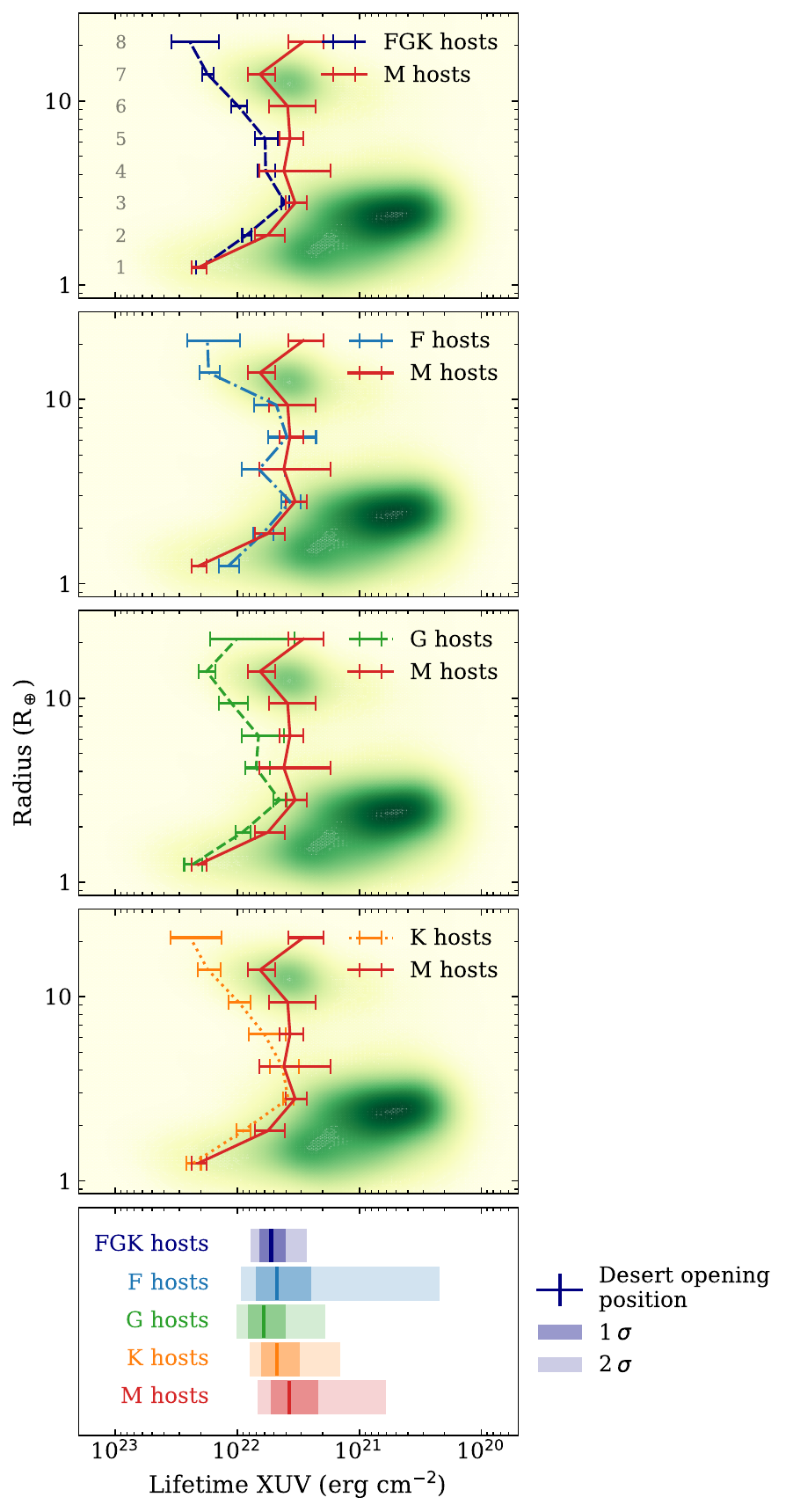}
    \caption{Comparison of the best fit desert boundaries in lifetime XUV space for M dwarf to the other samples we considered.}
    \label{fig:XUVboundaries}
\end{figure}

In Fig.~\ref{fig:XUVboundFit}, we show the results of our bootstrapping process for each sample in lifetime XUV space, analogous to Fig.~\ref{fig:sample-percentiles} for the other parameter spaces.

We note that in the background of Fig.~\ref{fig:XUVboundFit} are the individual systems plotted with specific lifetime XUV values. These are also quoted in Table~\ref{tab:sample}, where the values assume the median \textsc{Mors} track. The 16 and 84 percentiles were used for the errorbars, along with propagating through relevant uncertainties in age and semi-major axis. These values are provided to the reader for the purposes of reproducing or adapting our population-level analysis. We advise proceeding with \textit{significant} caution if using these lifetime XUV values of individual systems for other purposes. For assessing the high-energy irradiation of specific systems, observational constraints should be factored in wherever possible.

Some radius bins in Fig.~\ref{fig:XUVboundFit} have their lower 1 and/or 2-$\sigma$ percentiles (i.e. at the low-XUV end) somewhat unconstrained, particularly for the M hosted sample. This is likely due to two factors. Firstly, many of the larger radius bins have relatively few planets. Secondly, the large age uncertainties on many of the individual systems mean that a sizeable minority of draws in the bootstrap will tend towards very young ages, where the cumulative XUV up those drawn ages remains small. 

However, for the purposes of determining the desert boundaries, it is the high-XUV end we are interested in. Here, the errorbars on the averages of the upper 2-$\sigma$ percentiles are much better constrained across the full range of radius bins for each sample. This is likely reflective of the point we made above about how the cumulative XUV changes much more slowly above $\sim$Gyr, so even where age uncertainties are large, the resulting changes in the 97.7 percentile between lifetime XUV draws are relatively small.

To better compare the boundaries in lifetime XUV space across stellar types, in Fig.~\ref{fig:XUVboundaries} we plot the inferred boundary for the M hosted sample with the boundary for each other sample in turn. For small and intermediate radii, the boundaries in cumulative XUV space are very similar across all spectral types considered, with significant overlap of the plotted 1-$\sigma$ uncertainties in the boundary location. We also performed the same calculation for radius bins 3 to 5 as in \S\ref{sec:boundaries} to provide a quantitative value for the desert opening position in lifetime XUV space. We list these values in Table~\ref{tab:xuvsampleprops}, as well as plotting them in the bottom panel of Fig.~\ref{fig:XUVboundaries}. The agreement in boundary location across stellar types in this space is observational evidence that XUV-driven escape could indeed the primary mechanism forming the lower portion of the Neptunian desert. We discuss this further in \S\ref{sec:discuss}.

At larger radii, the boundaries in lifetime XUV space for the different samples deviate. This likely reflects the importance of other mechanisms such as HEM for forming and shaping the upper part of the desert. For the M dwarf hosted sample in particular, the very small sample size of giant planets also becomes relevant (as noted previously), with potential for the observed edge to change substantially as more such planets are discovered.

\section{Discussion}
\label{sec:discuss}

\subsection{Comparison with the HEM/tidal mechanism}
\label{ssec:HEMcompare}

\begin{figure}
    \centering
    \includegraphics[width=\columnwidth]{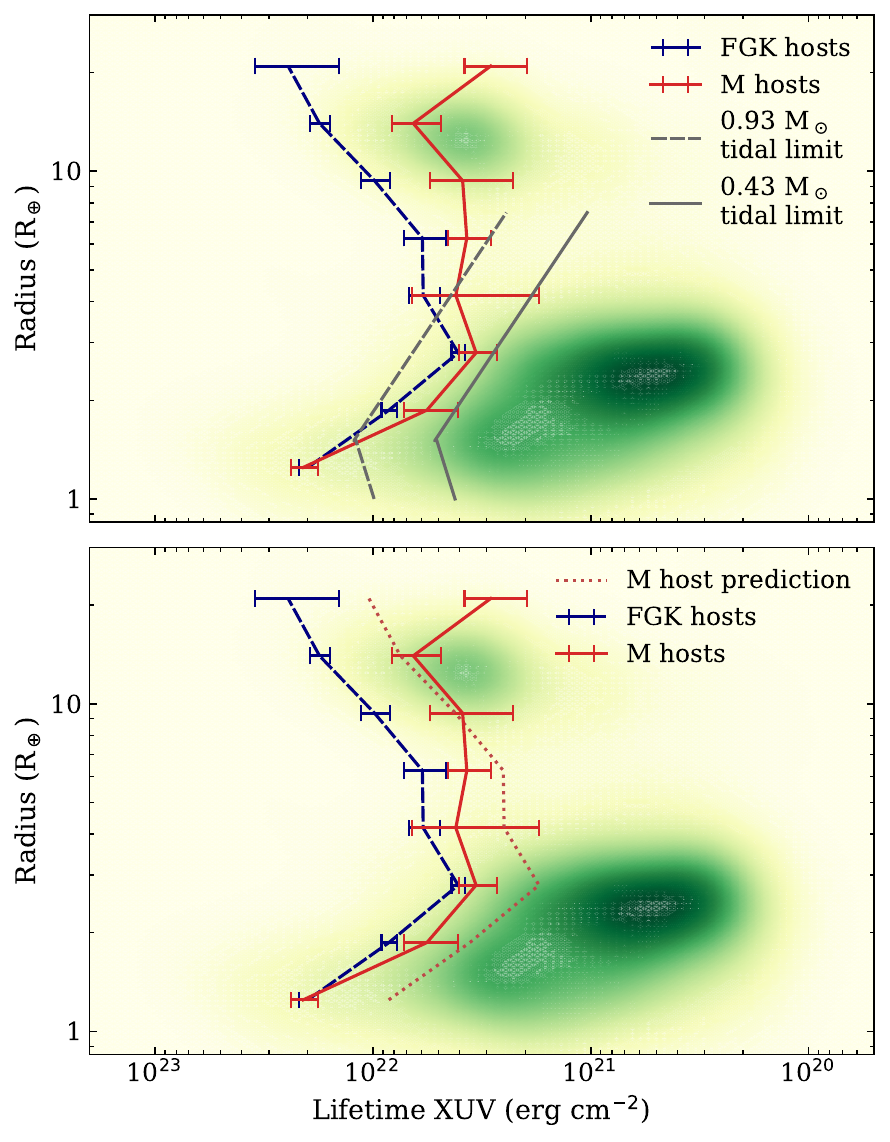}
    \caption{Reproductions of the comparison of the M and FGK combined samples in lifetime XUV space from the top panel of Fig.~\ref{fig:XUVboundaries}, with comparison to the HEM/tidal mechanism. In the top panel, we overplot tidal boundary curves corresponding to the average stellar mass of each sample (see main text for a description of how these were determined). In the bottom panel, we overplot an alternative HEM prediction for the M-hosted planets based on shifting the FGK boundary by the constant factor difference between the tidal lines in the top panel.}
    \label{fig:XUVboundaries_tidal}
\end{figure}

The range of planet radii over which there is evidence of high-eccentricity migration (HEM) contributing to forming the desert and nearby features has expanded recently. In one example, \citet{Thomas2026} found that among 4--8.5\,R$_\oplus$ planets, those in the desert and ridge were far less likely to have (small) planetary companions compared to those in the savanna (10 versus 70 per cent). This is in line with expectations for the HEM mechanism shaping the desert here, since most such companions would have been scattered away during the planet's inward migration \citep[][]{Mustill2015}. Notably, extending to even smaller radii, \citet{C-G2026} recently described being able to reproduce the desert boundary between 1.8 and 6\,R$_\oplus$ using the HEM mechanism.

In \S\ref{sec:XUVbound}, we suggested the overlap we find across stellar types in the lifetime XUV desert boundaries was indicative of XUV-driven escape shaping the lower part of the desert. Indeed, despite those recent studies demonstrating evidence for HEM dominating, our results across all parameter spaces we consider seem to fit with XUV-driven escape as high as 7.5\,R$_\oplus$ - our desert opening positions in Tables~\ref{tab:sampleprops} and~\ref{tab:xuvsampleprops} were calculated using bins 3 to 5 (i.e. 2.2 to 7.5\,R$_\oplus$), and appear to shift between stellar types as one would expect if XUV-driven escape were the dominant mechanism.

To more directly compare our results in lifetime XUV space with expectations from the tidal mechanism, we again replot the comparison of the FGK vs. M hosts in lifetime XUV space twice in Fig.~\ref{fig:XUVboundaries_tidal}. In the top panel, we overplot theoretical survival boundaries based on circularisation having to occur at a distance of at least twice the critical tidal radius, $r_{\rm tide}$, given by
\begin{equation}
    r_{\rm tide} = \eta R_{\rm p} \left( \frac{M_*}{M_{\rm p}}\right)^{\frac{1}{3}}.
    \label{eq:tidal}
\end{equation}
The constant $\eta$ depends on the internal structure of the planet, for which we adopt 4.5, the value with which \citet{C-G2026} reproduced the desert in period space from 1.8 to 6\,R$_\oplus$. However, we note that other values reported in the literature are typically lower, such as 2.7 \citep{Guillochon2011}, 2.44 \citep{Rappaport2013}, and 3.5 \citep{Matsakos2016}. To vary the planet radius, $R_{\rm p}$, and mass, $M_{\rm p}$, consistently, we adopted the mass-radius relation from \citet{Muller2024}. We then calculated the orbital distance threshold curves for 0.93 and 0.43\,M$_\odot$, corresponding to the average stellar masses of the FGK and M samples in Table~\ref{tab:xuvsampleprops}. Finally, we use the median \textsc{Mors} track for each stellar mass to find the cumulative XUV value at a typical field age of 5\,Gyr, and use this to scale the distance threshold curves into lifetime XUV space. We cut off plotting the resulting tidal boundary above radius bin 5 (i.e. above 7.5\,R$_\oplus$) in Fig.~\ref{fig:XUVboundaries_tidal}, as our mass radius relation becomes degenerate for giant planets, complicating our use of Eq.~\ref{eq:tidal}. This region is also not a primary focus of our study as there is existing consensus that the upper boundary is sculpted by HEM \citep[e.g.][]{Owen2018,Vissapragada2022}. It is only at smaller and intermediate radii where doubt remains as to the relative contributions of the evaporation and HEM mechanisms to forming the desert.

Comparing the two tidal limit lines, we observe that there should be a shift in the desert boundary to lower values of lifetime XUV for M hosted planets wherever the HEM mechanism dominates. While the theoretical curves are in roughly the right region of the plot, neither fit the observed boundaries in this space particularly well across all five bins. A key point to note though is that the absolute position and shape of the tidal limit curves are sensitive to e.g. the exact value of $\eta$ and mass-radius relation assumed. 

To investigate further, in the bottom panel of Fig.~\ref{fig:XUVboundaries_tidal} we take an alternative approach. The difference in lifetime XUV between the two tidal limit lines in the top panel is constant across radius, with a factor of 2.351 change. Since it is the relative change between FGK and M that we care about, in the bottom panel of Fig.~\ref{fig:XUVboundaries_tidal} we then plot the observed FGK boundary shifted by this factor, providing a prediction for where the M dwarf boundary should lie in this space if the tidal mechanism is dominant. Note that since the shift is constant across radius, we plot this prediction for the top three radius bins too.

A shift of the magnitude required to match HEM is not immediately obvious in the bottom panel of Fig.~\ref{fig:XUVboundaries_tidal}. Four of the first five radius bins for the M-hosted planets simultaneously overlap or lie closer to the FGK boundary than the prediction from HEM. However, as a statistical test, we calculate the $\sigma$ difference of the M dwarf boundary from both the FGK boundary and the prediction. For the prediction, we assumed equal fractional uncertainties on the prediction points to their corresponding FGK boundary points. Averaging across radius bins 3--5, the M dwarf boundary lies 0.78-$\sigma$ from the FGK boundary, and 0.85-$\sigma$ from the HEM prediction. 
Thus, while agreement with the FGK boundary is very slightly favoured over a shift to the expected HEM, our uncertainties are such that both scenarios are viable within 1-$\sigma$. Using bins 2--5 yields almost identical results, while including bin 1 leads to large errors on the averaged boundary because of the large increase in cumulative XUV for the boundary position for this bin. It is interesting to note though that the M dwarf boundary for bin 1 is very discrepant (4.2-$\sigma$) from its predicted value under HEM in the bottom panel of Fig.~\ref{fig:XUVboundaries_tidal}, favouring an XUV-driven origin. Additionally, bin 3 is 2.4-$\sigma$ discrepant from its HEM prediction in the same figure, while overlapping the observed FGK point.

One other finding, however, points towards a dominant contribution from XUV-driven escape across much of the lower desert. If one uses Eq.~\ref{eq:tidal} to find the tidal limit in orbital period space, another $M_*^{1/3}$ factor from Kepler's third law cancels out the existing term in the equation. As such, the boundary should be invariant with stellar mass in radius-period space. The M dwarf boundary for bins 3--5 is only 1.6-$\sigma$ discrepant from that for the combined FGK sample, but in splitting the latter up further Fig.~\ref{fig:sample-lines} shows an increasing shift to shorter orbital periods in the boundary from F, through G and K, to M. Further, radius bins 1, 2, 3, and 5 all individually show clear trends from F to M that cannot be explained by the measurement uncertainties. Interestingly, bin 4 shows significantly more overlap, with the F star boundary in particular shifting to shorter periods by $>1$\,d compared to bins 2, 3, 5, and 6, providing evidence of a possible contribution from HEM. However, in the lower boundaries of \citet{Cui2026spectype}, shown in the bottom panel of Fig.~\ref{fig:sample-lines}, there is a clear and consistent shift to shorter orbital periods for lower mass hosts all the way to at least 6\,R$_\oplus$.

Overall, the evidence points slightly in favour of XUV-driven escape shaping the lower boundary up to at least $\sim3.4$\,R$_\oplus$ (bins 1-3). Despite some ambiguities of our bin 4, the boundaries of our bin 5 and those in \citet{Cui2026spectype} suggest photoevaporation could act to sculpt the desert to potentially as high as 7.5\,R$_\oplus$. However, HEM cannot be fully ruled out, especially in lifetime XUV space. In the future, increasing the number of confirmed planets in the M-hosted sample in particular should help reduce the uncertainties and strengthen the evidence for either mechanism.

For the larger planets, only above $\sim7.5$\,R$_\oplus$ do we see clear diversion between FGK and M hosted planets in lifetime XUV space. The shift for M hosted planets is to lower XUV irradiation values, as expected from our theoretical tidal limits. Moreover, the prediction in the bottom panel of Fig.~\ref{fig:XUVboundaries_tidal} matches very well the position of radius bins 6 and 7. While Bin 8 is very discrepant from the prediction, there is only a single M-hosted planet in this bin, and so the accuracy of this point is questionable. These findings for giant planets are consistent with the long standing belief that the upper portion of the desert is sculpted by the HEM mechanism \citep[e.g.,][]{Owen2018,Vissapragada2022}.

\subsection{Comparison with previous studies}

\begin{figure}
    \centering
    \includegraphics[width=\columnwidth]{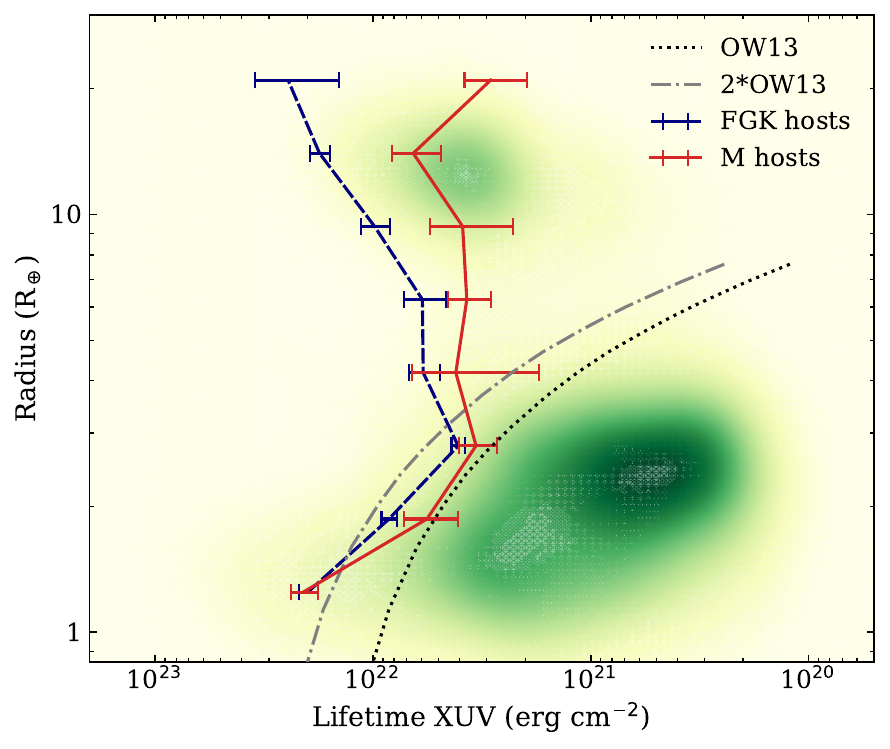}
    \caption{As Fig.~\ref{fig:XUVboundaries_tidal}, but with the \citet{Owen2013} theoretical evaporation threshold overplotted ("OW13"), as well as the same curve multiplied by two to account for that work only considering X-rays ("2*OW13").}
    \label{fig:XUVboundaries_OW13}
\end{figure}

Here, we compare our findings to several previous studies that touch on similar areas. 

First, \citet{McDonald2019} also found evidence of the desert boundary showing much less variation in lifetime XUV space than bolometric instellation space when considering different stellar host masses. Comparing our results to theirs, as a function of planetary radius, their desert boundary in lifetime XUV space shows substantial variation: from $8.23\times10^{20}$\,erg\,cm$^{-2}$ at $3.11-3.94$\,R$_\oplus$ to $1.42\times10^{22}$\,erg\,cm$^{-2}$ between 1.52 and 2.45\,R$_\oplus$. Our desert boundaries appear somewhat more stable across a wide radii range, varying by no more than a factor of 2--3 between 1.5 and 7.5\,R$_\oplus$. We do see a shift to much higher lifetime XUV dose in our lowest radius bin, although such planets are unlikely to retain any substantial envelope. It is worth noting when comparing the values directly (as opposed to the variation as a function of radius), \citet{McDonald2019} only consider X-ray flux and not total XUV.

In terms of approach, we used a more recent and comprehensive set of evolutionary tracks describing XUV emission evolution across stellar types, which can additionally account for differences in the initial rotation rate in stars of the same mass. These are an improvement over the simpler relations that the lookup table determined by \citet{McDonald2019} is based upon, and also push down to lower stellar mass (0.1 vs. 0.3\,M$_\odot$). Their increase in integrated X-ray with increasing stellar mass is largely below 0.6\,M$_\odot$, whereas we see the increase continuing to Solar-like stars and beyond. This is likely driven by the \textsc{Mors} tracks clustering much closer together in the saturation regime, where the \citet{Jackson2012} relations used by \citet{McDonald2019} saturate significantly below the canonical $\sim10^{-3}$ at higher stellar masses. The respective exclusion and inclusion of flares by \citet{Jackson2012} and \citetalias{Johnstone2021} might partially explain this difference, although one may expect this to also increase the observed saturation level of the low mass stars just as much as those of higher mass (at least with respect to the bolometric). In terms of the planet sample, we emphasise that our sample has a vastly greater number of not just M-hosted planets, but F-hosted planets too.

Second, in \citet{Hallatt2022} they hypothesise that the drop in occurrence of sub-Saturns ($4-8 R_{\oplus}$) toward shorter periods is a result of atmospheric mass loss processes. They define a ``survival core mass'', such that a sub-Saturn with a core at or above this mass is able to retain a radius of $> 4 R_{\oplus}$ at the end of its evolution, rather than being stripped below this into a sub-Neptune. This survival core mass grows with XUV exposure of a planet, and therefore decreases with orbital period and toward lower host star masses. 
Using evolution calculations including planet cooling, photoevaporation and bolometrically-driven mass loss, they measured the survival core mass as a function of orbital period and stellar mass. Assuming that the underlying core mass distribution of sub-Saturns is the same regardless of orbital period and stellar mass, they found a theoretical core mass distribution that reproduces the sub-Saturn occurrence rate profile (i.e. the fraction with cores above the survival mass) across orbital periods ${\lesssim}$100 days. Where the desert begins at $\sim3$ days around $\sim1\,M_{\odot}$ stars, this corresponds to an upper limit on sub-Saturn occurrence of ${\sim}$0.12 planets per 100 stars (measured per 0.3 dex P interval; see their Figure 1). They find this occurrence rate is reproduced by a survival core mass of $\sim 22-32\,M_{\oplus}$. They then estimate from their core mass distribution the corresponding orbital period around lower mass stars where this survival core mass is achieved, i.e., where the same occurrence rate should be reached, and thus where the boundary of the desert should be. This is at $\sim1.5$\,days for $0.8\,M_{\odot}$ stars, and $\sim 0.7$\,days for $0.5\,M_{\odot}$ stars.

Our average of bins 3-5 covers a similar range in radius to their definition of sub-Saturns, so we can compare our observationally calculated shift in the desert boundary to this prediction. We go from a $\sim 3.5$\,d opening for $\sim1\,M_{\odot}$ stars ($\sim1\sigma$ discrepant), to $\sim 2.9$\,d for $\sim0.8\,M_{\odot}$ stars, to $\sim 2.2$\,d for $\sim0.4\,M_{\odot}$ stars ($\sim3\sigma$ discrepant). Alternatively, we can frame this relatively, where our shift from 0.5 to $1\,M_{\odot}$ is 1.6x the period, and theirs is 4.3x the period. So while we see the expected behaviour of the shift to lower periods with decreasing stellar mass, we do not see it as dramatically as predicted by \citet{Hallatt2022}.  

\citetalias{Osborn2026} suggest the difference could be down to HEM also contributing to this boundary. However, we have two arguments against this in this work: first, that HEM predicts a boundary that is invariant in period when considering different stellar mass hosts, which we do not see; second, our boundary is consistent between stellar types in lifetime XUV flux at these intermediate (and low) radii, as would be expected if photoevaporation is the dominant mechanism in this regime. Our discrepancy with \citet{Hallatt2022} may instead stem partly from the difference in occurrence rate threshold defining the desert boundary, and potentially from the core mass distribution of hot sub-Saturns around M stars. This is more bottom-heavy than that of cool sub-Saturns around FGK stars, which was used to estimate the M star desert boundary \cite[such that hot sub-Saturn occurrence around M stars is lower than the estimate of][]{Hallatt2022}.

Third and finally, we compare to \citet{Owen2013}: in Fig.~\ref{fig:XUVboundaries_OW13}, we replot the top panel of Fig.~\ref{fig:XUVboundaries} but additionally overplot the theoretical evaporation threshold prediction from \citet[][see their figure 7]{Owen2013}. The lower portion of our observed boundary agrees reasonably well with their prediction, particularly when we multiply their curve by two in XUV space. This accounts for \citet{Owen2013} only considering X-ray emission, and assumes cumulative X-ray flux is equal to cumulative EUV flux (c.f. values in Table~\ref{tab:M+FGKcomp}). The prediction deviates away from the observed population at intermediate sizes, where the observed boundaries continue to agree across stellar types. This could result from e.g. a greater variation in core mass as a function of radius than \citet{Owen2013} assumes, or suggest that HEM does indeed contribute at these intermediate sizes. However, the latter would require an explanation for the observed shift in period space.

\subsection{Further inferences and notes on our assumptions}
\label{ssec:assumptions}

An implicit assumption in our original hypothesis (that the boundaries should be similar in lifetime XUV space wherever photoevaporation sculpts the desert) is that the composition of the close-in exoplanet population is similar across stellar types (where one is comparing between planets of commensurate radii). This is somewhat analogous to discussions around the radius valley, where a deeper valley implies a more homogenous core composition, while a wider variety of core compositions would dull the feature \citep[e.g.][]{Owen2017}. In that instance, the observational evidence so far points towards a relatively homogeneous core composition among super-Earths and sub-Neptunes \citep{VanEylen2018,Ho2023}, but this remains an open area of investigation, especially around low-mass stars \citep{Ho2024}. With considerable overlap between our boundaries at small and intermediate radii across our various samples, our results are suggestive that core compositional changes across stellar type are small among short orbital period planets ($\lesssim10$\,d).

Our hypothesis also includes assumptions about the XUV irradiation. We used single, all encompassing flux values for the entire XUV band. This assumes either 1) all wavelengths in this range contribute equally to the heating; or 2) XUV spectral changes with age are similar enough across stellar types that they can be considered a higher order effect within the size of our uncertainties. We know that scenario 1) is not the case in reality. Even before one considers preferential wavelengths due to specific atomic transitions, the absorption cross section of high-energy photons is inversely proportional to the photon energy. This means EUV photons are preferentially absorbed higher in the atmosphere.

Case 2) is more plausible. The surface X-ray flux emitted by a star, $F_{\rm X}$, and the average coronal temperature, $\overline{T}_{\rm cor}$, have been shown by \citet{Johnstone2015} to scale with each other according to
\begin{equation}
    \overline{T}_{\rm cor} \approx 0.11F_{\rm X}^{0.26}.
\end{equation}
$\overline{T}_{\rm cor}$ can be considered as a proxy for the general morphology of the X-ray spectrum. Back in \S\ref{ssec:XUVmodRes}, we saw that the rate of decline in XUV was similar across a wide range of stellar types, and so the rate of changes in the coronal plasma temperature should also be similar. However, the different X-ray luminosities and sizes of different stellar types mean the starting point ahead of those changes (i.e. the spectral morphology during the saturation regime) likely varies among them. Indeed, \citet{Johnstone2015} found that in the saturation regime, $\overline{T}_{\rm cor} \propto M_*^{0.6}$. Therefore, higher mass stars have hotter coronae initially and thus harder (higher energy dominated) X-ray spectra, and the changes in resulting mass loss rates could mean that the cumulative XUV required to strip a planet's envelope may vary with stellar mass. That all said, in \S\ref{ssec:compBoundwMors}, we found that the boundary shift between G and M stars in semi-major axis space scaled roughly linearly with the change in XUV flux. Together with our findings that the lifetime XUV boundaries are compatible across stellar types, this suggests that XUV spectral effects may indeed be smaller than our uncertainties can probe.

Our methodology in \S\ref{sec:XUVbound} also assumes that planets spend their entire lives at their drawn semi-major axis in each realisation, thereby ignoring migration effects. For those that underwent migration within the disc phase, the early-time deposition at their current semi-major axis makes this a fair assumption. However, those that arrive later will have undergone less cumulative XUV irradiation. This additionally reinforces our warning about using the quoted cumulative XUV values in Table~\ref{tab:sample} for individual system analyses.

\subsection{A word on the choice of parameter space}

It is common in the exoplanet literature to use (instantaneous) bolometric instellation space to investigate population features; indeed, we have done so here in order to compare to both other parameter spaces and previous studies. However, for the desert specifically, and in line with past studies such as \citet{McDonald2019}, we observe a strong dependence on the boundary location with stellar type in instellation space. This fits with the strong dependence of the cumulative XUV on stellar mass at fixed instellation in the bottom panel of Fig.~\ref{fig:cumulComb}, and should not be surprising. Rather, it is a simple reflection of a few properties: 1) $L_{\rm X}/L_{\rm bol}$ starts at a similar level across stellar types before declining; 2) the saturation timescale is a strong function of stellar mass; 3) $L_{\rm bol}$ is strongly correlated with stellar mass and roughly constant compared to the changes in $L_{\rm X}$; and 4) under XUV-driven escape, the bolometric flux of the star is irrelevant to the mass loss rate. Instellation space is therefore not very informative for investigating the desert, and we recommend it should never be the only parameter space considered by individual studies.

What parameter space(s) should be used instead? As it is the most relevant quantity, lifetime-integrated XUV is the gold standard for investigating the contribution of photoevaporation towards carving the desert. However, the full stellar XUV history for individual stars is not directly observable, and so using this space requires careful treatment. Orbital period is perhaps the best choice among direct observables, particularly when using a sample encompassing a wide range of host stellar types. The desert boundaries appear to vary over a much smaller range with stellar type compared to semi-major axis or bolometric instellation, while also having the benefit that the tidal limit for the HEM mechanism being expected to be invariant as a function of stellar mass, save for planetary mass-radius relation changes. Therefore, any small observed shifts, as seen in our study and \citet{Cui2026spectype}, can be more confidently attributed to differences in cumulative XUV, and thus the photoevaporation mechanism.

\subsection{Future outlook}

There are several promising avenues for improving further upon the work we have presented. In some sense, planet mass would make for a more natural y-axis than planet radius. Mass-radius relations such as the one we employed from \citet{Muller2024} in \S\ref{ssec:HEMcompare} tend to be degenerate in radius, but not in mass. However, due to current technology and the ability to monitor many thousands of stars simultaneously, planets are generally much easier to discover and detect with transits rather than radial velocity signals, especially for smaller/less massive planets. Our sample of well-characterised M dwarfs hosting planets useful for the desert work in this paper is already quite small; requiring a precise mass measurement would reduce the sample size further.

While \textit{TESS} continues to increase our sample sizes, the impending launch of \textit{PLATO} will further expand our known, well-characterised population. For the M stars especially, the increase in detected systems over the next few years should reduce the size of the uncertainties substantially. Investigations of the M dwarf desert using planetary mass rather than radius should also be possible thanks to dedicated ground-based precision radial velocity follow up built into the mission. For F and G stars, alongside the increasing the known planet population, \textit{PLATO} will have the added benefit of measuring precise stellar parameters using asteroseismology. As well as stellar masses and radii, this will provide precise stellar ages for potentially several thousand main sequence stars \citep{Goupil2024}. This will allow a much-improved assessment of XUV time evolution of main sequence stars, which beyond 1\,Gyr relies largely on extrapolation due to the lack of precise age measurements \citep[e.g.][]{Jackson2012}. Moreover, there is the hope of tracking features like the desert in the planet population as a function of age. However, for K- and especially M-hosted systems, the mission's impact on planet demographics will almost solely be through increases in sample sizes from new discoveries, rather than additionally benefitting from precise, asteroseismic stellar parameters.

\section{Conclusions}

The boundaries of the Neptunian desert vary significantly depending not just on which parameter space one considers, but also on the mass of the host star. Increases in the number of well-characterised M-hosted planets have allowed us to stretch the mass range over which stellar type subsamples can be separately assessed. 

For the desert boundaries at small and intermediate radii, with decreasing stellar mass we observe: 1) a gentle shift to shorter periods, matching \citet{Cui2026spectype}; 2) a stronger shift to shorter orbital distances; and 3) a very large shift to lower instantaneous bolometric fluxes.

For comparison with the XUV-driven escape mechanism, we used a widely-employed semi-empirical model of XUV emission as a function of stellar mass and age. We emphasise that M dwarfs emit less XUV energy per unit time than Solar-like stars, and it is only as a fraction of their bolometric emission that they remain ``more active" than their more massive counterparts. This is a key nuance that is not always captured in the exoplanet literature. Integrating to a typical field age of 5\,Gyr, we see that the behaviour of cumulative XUV flux as a function of stellar mass at fixed values of semi-major axis, orbital period, and bolometric instellation matches the observed desert boundary shifts. These findings bolster the argument that the photoevaporation mechanism dominates the lower portion of the desert. 

We also examined the desert boundaries in cumulative XUV flux space. We find divergence between the FGK and M samples above 7.5\,R$_\oplus$, consistent with expectations from HEM. Below this radius, we find considerable overlap across stellar types. Under the assumption that the composition of planets of a given radius is the same or very similar across stellar types F through M, this overlap again points to the XUV-driven escape mechanism shaping the desert. However, a boundary shift to lower lifetime XUV fluxes for lower mass stars of the magnitude expected under the HEM mechanism cannot be ruled out, and requires further examination as the M-hosted sample expands.

Such expansion is expected in the coming years through new and continuing discoveries with \textit{PLATO} and \textit{TESS}. This will enable substantial improvements in our boundary measurement uncertainties. Increases in the number of M-hosted systems with well-characterised planet masses will enable the performance of a similar analysis using masses rather than radii. The combination of these should help further distinguish between the XUV-driven and HEM mechanisms, particularly at intermediate radii.

\section*{Acknowledgements}

We thank the anonymous referee for their comments which helped improve the clarity of this manuscript. This research was in part funded by the UKRI (Grants ST/X001121/1, EP/X027562/1). This research has made use of the NASA Exoplanet Archive, which is operated by the California Institute of Technology, under contract with the National Aeronautics and Space Administration under the Exoplanet Exploration Program.

\section*{Data Availability}


The data used for determining the desert boundaries are provided in Appendix~\ref{app:sample}. The Mors code used for generating lifetime XUV tracks available at \url{https://github.com/colinphilipjohnstone/Mors}.



\bibliographystyle{mnras}
\bibliography{desert_paper} 




\clearpage

\appendix

\onecolumn
\section{Planet sample for Neptunian desert boundary study}
\label{app:sample}
\FloatBarrier

\begin{figure*}
    \centering
    \includegraphics[width=\linewidth]{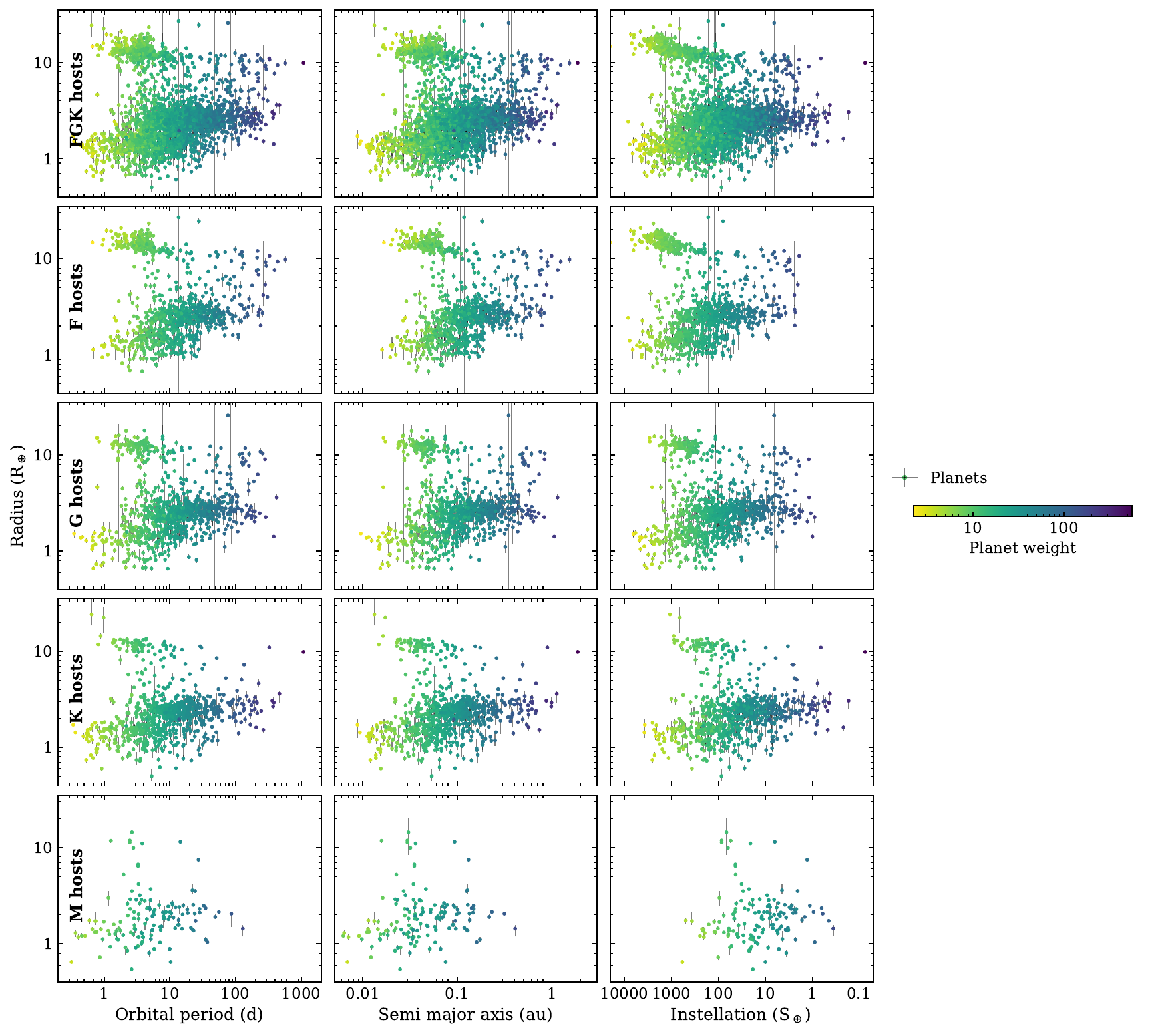}
    \caption{The sample used for the study of the Neptunian desert boundaries, described in \S\ref{sec:boundaries} and Table~\ref{tab:sampleprops}, and provided in full in Table~\ref{tab:sample}. Only planets with transit ${\rm S/N} > 7.1$ are shown.}
    \label{fig:sample}
\end{figure*}

\clearpage
\begin{landscape}

\begin{table*}
\centering
\small
\caption{Sample used for the study of the Neptunian desert boundaries.}
\label{tab:sample}
\begin{threeparttable}
\begin{tabular}{lllllllllllllllllllll}
\toprule
Planet & $a$ & $T_{\rm eff}$ & $R_{\star}$ & $M_{\star}$ & $L$  & $S$ & $\tau_{\star}$& $P_{\rm orb}$ & $R_{p}/R_{\star}$ & $R_{p}$ & $T_{\rm dur}$ & $\delta$ & $N$ & $N_{S}$ & $\sigma_{\rm CDPP}$ & S/N & $P_{\rm trans}^{-1}$ & $P_{\rm det}^{-1}$ & $w$ & $\int F_{\rm XUV, p}$ \\
       & (au) & (K) & ($R_{\odot}$) & ($M_{\odot}$) & ($L_{\odot}$) & ($S_{\oplus}$) & (Gyr) & (days) & & ($R_{\oplus}$) & (hrs) & & & & & & & & & (erg\,cm$^{-2}$) \\
\midrule
K00752.01	&	0.086	&	5610	&	0.90	&	0.94	&	0.73	&	99	    &	3.65	&	9.49	&	0.022344	&	2.20	&	3.59	&	0.00050	&	142	&		&	207.29	&	287.63	&	20.43	&	1.03	&	20.99	&	8.78E+20	\\
K00752.02	&	0.275	&	5610	&	0.90	&	0.94	&	0.73	&	10	    &	3.65	&	54.42	&	0.027954	&	2.76	&	5.36	&	0.00078	&	25	&		&	184.96	&	211.70	&	65.46	&	1.03	&	67.27	&	8.55E+19	\\
K00755.01	&	0.036	&	5660	&	0.92	&	0.94	&	0.79	&	630	    &	4.61	&	2.53	&	0.024064	&	2.42	&	1.74	&	0.00058	&	515	&		&	325.80	&	404.22	&	8.28	&	1.03	&	8.51	&	5.52E+21	\\
K00756.01	&	0.098	&	5837	&	1.03	&	1.03	&	1.12	&	116	    &	3.18	&	11.09	&	0.036779	&	4.14	&	3.66	&	0.00136	&	95	&		&	235.39	&	561.33	&	20.53	&	1.03	&	21.10	&	7.89E+20	\\
K00756.02	&	0.051	&	5837	&	1.03	&	1.03	&	1.12	&	431	    &	3.18	&	4.13	&	0.026133	&	2.94	&	2.04	&	0.00068	&	240	&		&	299.06	&	354.53	&	10.63	&	1.03	&	10.93	&	2.94E+21	\\
\multicolumn{21}{c}{\vdots} \\
toi892.01	&	0.100	&	6163	&	1.33	&	1.19	&	2.33	&	231	    &	2.72	&	10.63	&	0.079	    &	11.48	&	5.42	&	7530.0	&	2	&	1	&	220.61	&	51.60	&	16.19	&	1.03	&	16.64	&	7.94E+20	\\
toi905.01	&	0.046	&	5434	&	0.93	&	0.95	&	0.69	&	319	    &	4.96	&	3.74	&	0.1311	    &	13.30	&	2.04	&	15256.9	&	42	&	6	&	1943.18	&	332.65	&	10.73	&	1.03	&	11.02	&	3.36E+21	\\
toi942.01	&	0.049	&	5043	&	0.83	&	0.83	&	0.40	&	169	    &	10.30	&	4.32	&	0.03995	    &	3.61	&	3.18	&	1480.0	&	6	&	1	&	556.19	&	6.55	&	12.71	&	10.09	&	128.16	&	2.80E+21	\\
toi954.01	&	0.052	&	5932	&	1.78	&	1.35	&	3.62	&	1361	&	3.40	&	3.68	&	0.0462	    &	8.98	&	4.54	&	2070.0	&	7	&	1	&	178.90	&	30.81	&	6.22	&	1.03	&	6.39	&	0	        \\
toi966.01	&	0.046	&	6231	&	1.19	&	1.15	&	1.93	&	892	    &	2.07	&	3.41	&	0.11091	    &	14.34	&	3.21	&	16294.4	&	22	&	3	&	1748.80	&	208.27	&	8.43	&	1.03	&	8.66	&	3.40E+21	\\
\bottomrule
\end{tabular}
\begin{tablenotes}
\item The sample is described in \S\ref{sec:boundaries} and depicted in Fig.~\ref{fig:sample}, and is an updated version of the sample presented in Table E.1 of \citetalias{Osborn2026}. The columns have the same definitions as given there, and so we refer the reader to the table notes and text there for a full explanation of the table contents. 
\item There is, however, the addition of two columns from previous: the stellar age, $\tau_{\star}$, taken from \citet{Berger2023}; and the lifetime XUV flux experienced by the planet, $\int F_{\rm XUV, p}$, the calculation of which is described in \S\ref{sec:XUVbound}. Where this value is zero, the stellar mass is larger than bounds allow for calculation of the lifetime XUV flux in Mors (which can model 0.1 to 1.2\,M$_\odot$). 
\item We note that this table is truncated (error columns are omitted and precisions are reduced); the full version can be found on the Strasbourg astronomical Data Center (CDS).
\end{tablenotes}
\end{threeparttable}
\end{table*}

\end{landscape}

\clearpage
\twocolumn

\section{Separate X-ray and EUV time evolutions}
\label{app:timeEvo}
\FloatBarrier

\begin{figure}
    \centering
    \includegraphics[width=\columnwidth]{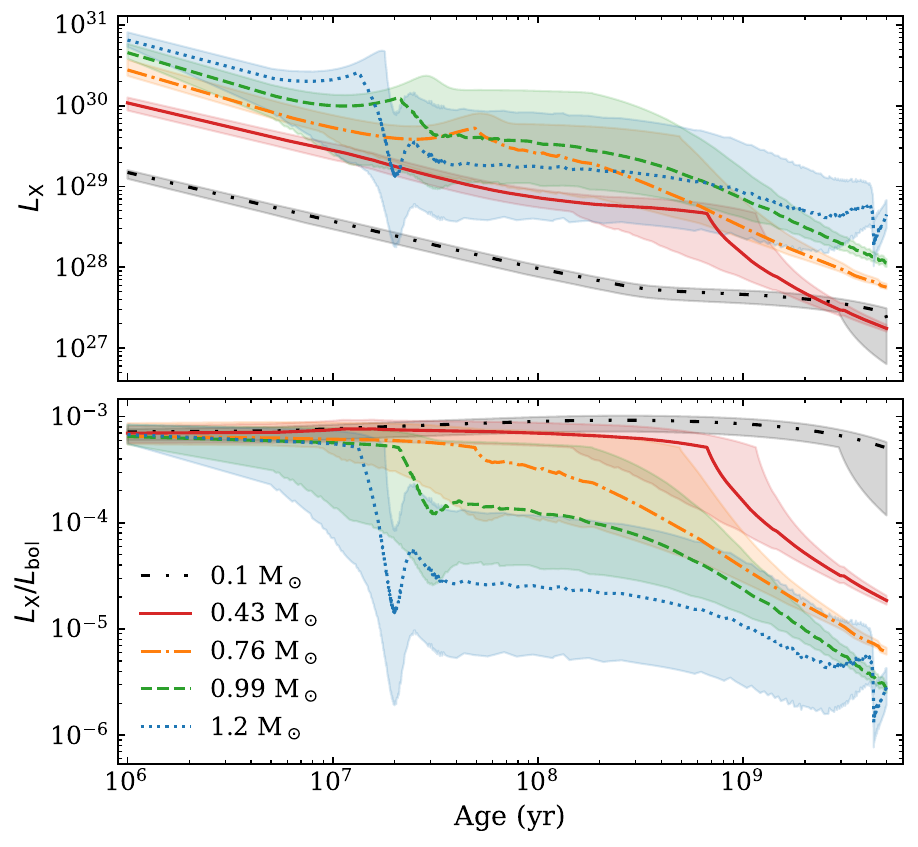}
    \caption{As Fig.~\ref{fig:Lxuv+ratios} but for the X-ray band only.}
    \label{fig:Lx+ratios}
\end{figure}

\begin{figure}
    \centering
    \includegraphics[width=\columnwidth]{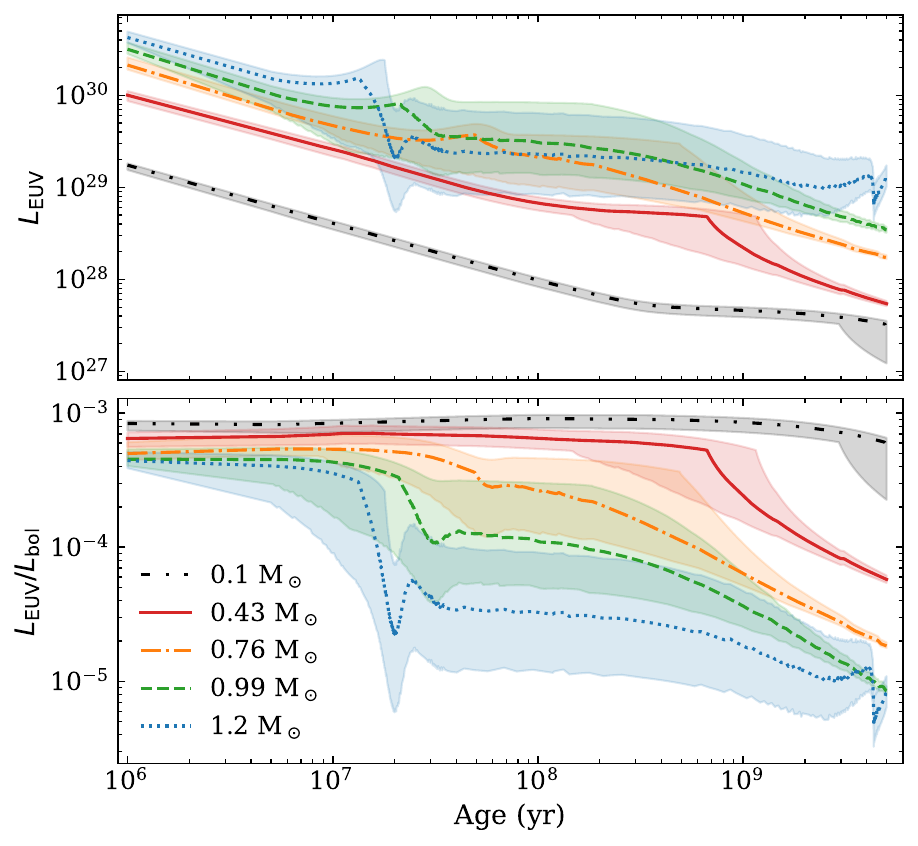}
    \caption{As Fig.~\ref{fig:Lxuv+ratios} but for the EUV band only.}
    \label{fig:Leuv+ratios}
\end{figure}

In Fig.~\ref{fig:Lxuv+ratios} we presented the time evolution of several stars for the total XUV band. For comparison, Figures \ref{fig:Lx+ratios} and \ref{fig:Leuv+ratios} show the separate X-ray and EUV contributions as a function of age. While the EUV tracks are very slightly more fanned out at the earliest ages, the overall shape remains consistent between X-ray and EUV, motivating our plotting only of the total XUV energy range in the main figures of the paper.


\bsp	
\label{lastpage}
\end{document}